\documentclass[%
 aip,
 amsmath,amssymb,
 reprint,%
 citeautoscript,
]{revtex4-2}

\usepackage{amsbsy}
\usepackage{graphicx}
\usepackage{dcolumn}
\usepackage{bm}
\usepackage{multirow}
\usepackage[dvipsnames]{xcolor}
\usepackage{chemformula}
\usepackage{gensymb}
\usepackage{amsmath}
\usepackage[utf8]{inputenc}
\usepackage[T1]{fontenc}
\usepackage{mathptmx}
\usepackage{siunitx}
\usepackage{cancel}

\newlength{\CapLen}
\AtBeginDocument{\settoheight{\CapLen}{K}}
\newcommand{\Kappa}{\resizebox{!}{\CapLen}{$\kappa$}}
\newcommand\footnoteref[1]{\protected@xdef\@thefnmark{\ref{#1}}\@footnotemark}

\begin{document}
\title{Molecular Dipole Moment Learning via Rotationally Equivariant Gaussian Process Regression with Derivatives in Molecular-orbital-based Machine Learning}
\author{Jiace Sun}
\affiliation{Division of Chemistry and Chemical Engineering, California Institute of Technology, Pasadena, CA 91125, USA}
\author{Lixue Cheng}
\affiliation{Division of Chemistry and Chemical Engineering, California Institute of Technology, Pasadena, CA 91125, USA}
\author{Thomas F. Miller III}
\email{tfm@caltech.edu.}
\affiliation{%
Division of Chemistry and Chemical Engineering, California Institute of Technology, Pasadena, CA 91125, USA
}%

\date{\today}
\begin{abstract}
This study extends the accurate and transferable molecular-orbital-based machine learning (MOB-ML) approach to modeling the contribution of electron correlation to dipole moments at the cost of Hartree--Fock computations. 
A molecular-orbital-based (MOB) pairwise decomposition of the correlation part of the dipole moment is applied, and these pair dipole moments could be further regressed as a universal function of molecular orbitals (MOs).
The dipole MOB features consist of the energy MOB features and their responses to electric fields. 
An interpretable and rotationally equivariant Gaussian process regression (GPR) with derivatives algorithm is introduced to learn the dipole moment more efficiently. 
The proposed problem setup, feature design, and ML algorithm are shown to provide highly-accurate models for both dipole moment and energies on water and fourteen small molecules. 
To demonstrate the ability of MOB-ML to function as generalized density-matrix functionals for molecular dipole moments and energies of organic molecules, we further apply the proposed MOB-ML approach to train and test the molecules from the QM9 dataset. 
The application of local scalable GPR with Gaussian mixture model unsupervised clustering (GMM/GPR) scales up MOB-ML to a large-data regime while retaining the prediction accuracy. 
In addition, compared with literature results, MOB-ML provides the best test MAEs of 4.21 mDebye and 0.045 kcal/mol for dipole moment and energy models, respectively, when training on 110000 QM9 molecules. The excellent transferability of the resulting QM9 models is also illustrated by the accurate predictions for four different series of peptides.
\end{abstract}

\maketitle

\section{Introduction}
\label{introduation}
Applications of machine learning (ML) to electronic structure theory have grown rapidly with an increasing number of studies in a variety of chemical systems and applications ~\cite{von2020retrospective,westermayr2021perspective}, such as directly predicting the molecular properties, developing force fields and interatomic potentials, and designing novel and efficient catalysts~\cite{freeze2019search,elton2019deep}, drugs~\cite{von2005variational,von2007alchemical,Popova2018,yang2019concepts} and materials~\cite{oganov2019structure,tran2020multi}. Important applications of machine learning include predicting chemical properties directly to reduce computational costs via supervised learning~\cite{Grisafi2018,manzhos2020neural}, detecting the patterns of chemical spaces via unsupervised learning~\cite{ceriotti2019unsupervised,tshitoyan2019unsupervised}, and proposing more suitable chemical systems via reinforcement learning ~\cite{Popova2018,zhou2019optimization} and generative models ~\cite{Sanchez-Lengeling2018,schwalbe2020generative}. 

Numerous approaches have been presented in the field of machine learning for electronic structure during the last decades to aid in the learning of molecular energies and other molecular properties ~\cite{Bartok2010,rupp2012fast,VonLilienfeld2013,hansen2013assessment,Ceriotti2014,ramakrishnan2015big,Tuckerman,kearnes2016molecular,Paesani2016, Behler2016,schutt2017quantum,schutt2017schnet,Smith2017,Welborn2018,wu2018moleculenet,Nguyen2018,Yao2018, Fujikake2018,unke2019physnet,Cheng2019,cheng2019regression,dick2020machine,deephf,qiao2020orbnet,qiao2020multi,hermann2020deep,deringer2021origins,christensen2021orbnet,husch2020improved,lee2020analytical,sun2021molecular,Karandashev2022,veit2020predicting,orbnetequi,klicpera2020fast,liu2022spherical,painn,faber2018alchemical,huang2020quantum,bartok2017machine,OQML}.
Molecular-orbital-based machine learning (MOB-ML)\cite{Welborn2018,Cheng2019,cheng2019regression,husch2020improved,lee2020analytical,sun2021molecular,cheng2022} is one such method that uses molecular orbital (MO) information from Hartree--Fock (HF) computation to create a simpler and more direct mapping from the MO-based (MOB) features to the correlation energies. 
By utilizing quantum level information as features, MOB-ML has shown great efficiency and transferability for highly accurate predicted molecular energies with few training data. The introduction of a general regression with a clustering learning framework and a scalable Gaussian process algorithm further scales up MOB-ML to learn a large amount of data.\cite{sun2021molecular,cheng2022} 

The responses of energy to some external variables in the time-independent framework are frequently studied molecular properties \cite{gonze1995adiabatic,LRTDDFT} in ML for quantum chemistry, such as atomic force, electric or magnetic dipole moments, and electric or magnetic polarizability.
The electric dipole moment (referred as dipole), defined as the first-order response of the molecular energy to an external electric field, is an essential component in computational spectroscopy. \cite{grunenberg2011computational}.
However, it remains challenging to efficiently and accurately compute dipole moments with density functional theory. \cite{hait2018accurate, LRTDDFT}
Several previous efforts have been made to model the dipole moment and other response properties via ML methods 
\cite{darley2008beyond,schutt2017schnet,OQML,orbnetequi,veit2020predicting,painn,liu2022spherical,klicpera2020fast,unke2019physnet,wilkins2019accurate}. 
MOB-ML also shows a great potential to model any molecular properties determined by electronic wavefunctions, including these response properties. For instance, highly-accurate atomic force predictions could be directly provided by the MOB-ML models trained on energy labels only. \cite{lee2020analytical}

In this work, we consider the application of MOB-ML to learn general time-independent linear response properties using dipole moments as an example. 
An accurate and transferable MOB-ML framework is introduced to learn the contribution of electron correlation to the dipole moment as a summation of pairwise contributions. 
These pair dipole moments can be learnt as the functions of MOB features and their derivatives to electric fields via a rotationally equivariant Gaussian process regression (GPR) algorithm, i.e., GPR with derivatives. \cite{GPbook}
We also adapt the local GPR with the Gaussian mixture model (GMM) unsupervised clustering (GMM/GPR) algorithm \cite{cheng2022} to scale up the training and reduce the training costs without sacrificing accuracy and transferability. \cite{cheng2022accurate}
The accuracy and efficiency of the proposed MOB-ML approach are tested on various benchmark systems, including water, fourteen small molecules, the QM9 benchmark dataset\cite{ramakrishnan2014quantum}, and four series of peptides\cite{veit2020predicting}. The literature methods to predict dipole moments and energies for QM9 datasets are also included for comparison.

\section{Theory and Methods}
\subsection{Review of MOB-ML to learn molecular energies}
MOB-ML is motivated by the Nesbet theorem \cite{Nesbet1958} to learn the pairwise contribution of the correlation energy $E^\text{corr}$ as a universal function of occupied molecular orbitals (MO) represented by features computed from Hartree--Fock (HF):
\begin{align}
     E^\text{corr} &= \sum_{ij\in\text{occ}} \epsilon_{ij},\label{eq:energy_pair_decomposition}\\
     \epsilon_{ij} &\approx \epsilon^{\text{ML}}[\boldsymbol{f}^\epsilon_{ij}({\phi_k})] \label{eq:energy_universal},
\end{align} 
where $\epsilon_{ij}$ is the pair energy corresponding to the occupied MO $i$ and $j$, $\boldsymbol{f}^\epsilon_{ij}$ is the MOB-ML feature vector for the MO pair $i$ and $j$, and ${\phi_k}$ is the set of all the MOs. 
The symmetrization of $i, j$ orbitals is also adapted to ensure $\boldsymbol{f}^\epsilon_{ij}=\boldsymbol{f}^\epsilon_{ji}$. 
The feature vector is consisted of Fock matrix elements $F_{pq}$ and part of the electron repulsion integrals $[\Kappa^{pq}]_{mn} = \langle pq | mn \rangle$ \cite{Cheng2019,husch2020improved}:
\begin{equation} \label{eq:energy_feature}
\begin{split}
    \boldsymbol{f}_{ij} =& \{ \{F_{ii}, F_{ij}, F_{jj}\}, \{F_{ik}\}, \{F_{jk}\}, \{F_{ab}\}, \\
    &   \{[\Kappa^{ii}]_{ii}, [\Kappa^{ii}]_{jj}, [\Kappa^{jj}]_{jj}\}, \{[\Kappa^{ii}]_{kk}\}, \{[\Kappa^{jj}]_{kk}\}, 
    \\
    & \{[\Kappa^{ii}]_{aa}\}, \{[\Kappa^{jj}]_{aa}\}, \{[\Kappa^{aa}]_{bb}\}, \{[\Kappa^{ij}]_{ij}\}, \{[\Kappa^{ik}]_{ik}\}, 
    \\
    &\{[\Kappa^{jk}]_{jk}\}, \{[\Kappa^{ia}]_{ia}\}, \{[\Kappa^{ja}]_{ja}\}, \{[\Kappa^{ab}]_{ab}\} \}.
\end{split}
\end{equation}
which is referred as "energy feature set" in this study.
The energy feature generation and sorting approach is the same as in Ref. \citenum{husch2020improved}. The reference pair energies can be calculated in most of post-HF methods, such as second-order Møller-Plesset perturbation theory \cite{LMP2} (MP2) and various coupled cluster theories, including CCSD and CCSD(T) \cite{LCCSD,LCCSDT}.

\subsection{MO Decomposition of dipole moments in MOB-ML} \label{sec:decomposition}

\begin{figure}[htbp]
    \centering
    \includegraphics[width=0.8\columnwidth]{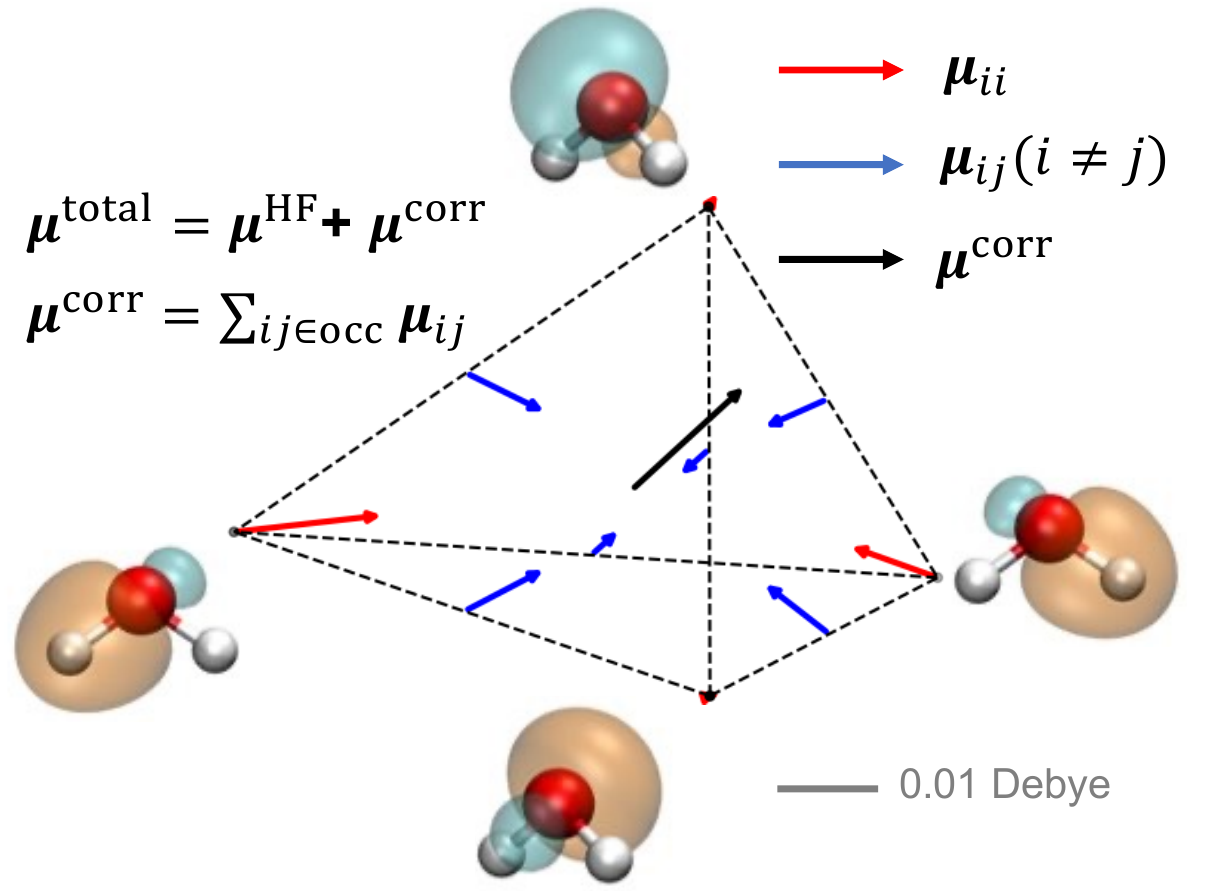}
    \caption{Example decomposition of dipole moments as a sum of pairwise MO contributions for a water molecule. The four vertices of the tetrahedron represent the self-interactions of four MOs (indexing as $ii$), and the six edges connecting the vertices represent the interactions between two MOs (indexing as $ij$). The pair dipoles $\boldsymbol{\mu}_{ii}$ and $\boldsymbol{\mu}_{ij}$ are shown in red and blue arrows with correct direction and scaling, respectively. Four MOs are also shown next to the corresponding vertex $i$. The relative length scaling is also shown using a grey line.}
    \label{fig:water_dipole}
\end{figure}

For a given system, the electric dipole moments $\boldsymbol{\mu}$ can be expressed as the linear response of the energy $E$ with respect to external electric field $\pmb{\mathcal{E}}$, i.e.,
\begin{equation}
    \boldsymbol{\mu} = \nabla_{\pmb{\mathcal{E}}} E,
\end{equation}
which could be further expressed as the sum of HF and correlation components:
\begin{equation}
    \boldsymbol{\mu}=\nabla_{\pmb{\mathcal{E}}} E^{\text{HF}} + \nabla_{\pmb{\mathcal{E}}} E^{\text{corr}} = \boldsymbol{\mu}^{\text{HF}} + \boldsymbol{\mu}^{\text{corr}}
\end{equation}
The correlation part $\boldsymbol{\mu}^{\text{corr}}$ can then be decomposed on pairs of occupied orbitals similar to Eq.~\ref{eq:energy_pair_decomposition}:
\begin{equation}
    \boldsymbol{\mu}^{\text{corr}} = \nabla_{\pmb{\mathcal{E}}} E^{\text{corr}} = \sum_{ij\in\text{occ}} \nabla_{\pmb{\mathcal{E}}} \epsilon_{ij} = \sum_{ij\in\text{occ}} \boldsymbol{\mu}_{ij}.
\end{equation}
$\boldsymbol{\mu}_{ij}$ is referred to as pair dipole moments and is regressed by ML similar to Eq.~\ref{eq:energy_universal}. Compared with Eq.~\ref{eq:energy_universal}, we add the feature derivatives information $\nabla_{\pmb{\mathcal{E}}} \boldsymbol{f}_{ij}({\phi_k})$ as part of the features motivated by $\boldsymbol{\mu}_{ij}=\nabla_{\pmb{\mathcal{E}}} \epsilon_{ij}$.
\begin{equation} \label{eq:response_universal}
    \boldsymbol{\mu}_{ij} \approx  \boldsymbol{\mu}^{\text{ML}}[\boldsymbol{f}_{ij}({\phi_k}), \nabla_{\pmb{\mathcal{E}}} \boldsymbol{f}_{ij}({\phi_k})]. 
\end{equation}
Figure \ref{fig:water_dipole} displays an example of the dipole moment decomposition on a water molecule to facilitate an understanding of this decomposition.

\subsection{Feature design of dipole learning in MOB-ML}
\label{sec:dipole_feature}
The energy feature set $\boldsymbol{f}^\epsilon_{ij}$ (Eq.~\ref{eq:energy_feature}) includes enough information to model molecular energy and satisfies different invariance properties, including translational, rotational, and orbital permutational invariances. \cite{Cheng2019,husch2020improved} 
To efficiently model the dipole moments, we additionally include the responses of feature vector to electric field $\pmb{\mathcal{E}}$, i.e., $\nabla_{\pmb{\mathcal{E}}}\boldsymbol{f}^\epsilon_{ij}$, in the design of dipole feature set $\boldsymbol{f}^{\boldsymbol{\mu}}_{ij}$ .
\begin{equation} \label{eq:dipole_featue}
    \boldsymbol{f}^{\boldsymbol{\mu}}_{ij} = 
    \{\boldsymbol{f}^\epsilon_{ij}, 
    \nabla_{\pmb{\mathcal{E}}} \boldsymbol{f}^\epsilon_{ij} \}
\end{equation}
However, the direct definition of $\nabla_{\pmb{\mathcal{E}}}\boldsymbol{f}^\epsilon_{ij}$ does not satisfy translational invariance due to the dependence of the Fock matrix elements $\nabla_{\pmb{\mathcal{E}}} F_{pq}$ on the positional operator matrix element $\boldsymbol{r}_{pq}$.

\begin{align}
    F_{pq} &= h_{pq} + \sum_{k=1}^{n} (2J^k_{pq} - K^k_{pq}) + \cancel{\boldsymbol{r}_{pq} \cdot \pmb{\mathcal{E}}},
    \\
    \nabla_{\pmb{\mathcal{E}}} F_{pq} \Big|_{\pmb{\mathcal{E}}=0} &= \nabla_{\pmb{\mathcal{E}}} \left( h_{pq} + \sum_{k=1}^{n} (2 J^k_{pq} - K^k_{pq}) \right) \Big|_{\pmb{\mathcal{E}}=0} \nonumber \\
    &+  \cancel{\nabla_{\pmb{\mathcal{E}}} \boldsymbol{r}_{pq} \cdot \pmb{\mathcal{E}} \Big|_{\pmb{\mathcal{E}}=0}} + \bcancel{\boldsymbol{r}_{pq}}  \label{eq:fock_derivative}     
\end{align}
where $h_{pq}$ are the one-electron Hamiltonian matrix elements, $J^k_{pq}$ and $K^k_{pq}$ are Coloumb and exchange matrix elements of the $k$th orbital, and $\boldsymbol{r_{pq}}$ are the position operator matrix elements. 
The $\boldsymbol{r}_{pq} \cdot \pmb{\mathcal{E}}$ and  $\nabla_{\pmb{\mathcal{E}}} \boldsymbol{r}_{pq} \cdot \pmb{\mathcal{E}}$ terms vanish when $\pmb{\mathcal{E}}=\boldsymbol{0}$. However, the existence of $\boldsymbol{r}_{pq}$ term in Eq.~\ref{eq:fock_derivative} results in a non-translataional invariant dipole feature design. 
Therefore, a redefinition of the derivatives of Fock matrix elements is adapted by subtracting the $\boldsymbol{r}_{pq}$ term to make dipole feature vector (Eq.~\ref{eq:dipole_featue}) translataional invariant.

\subsection{Review of scalable GPR algorithms in MOB-ML}
In MOB-ML, Gaussian process regression (GPR) is used to model the molecular energies with high accuracy and great transferability. \cite{Welborn2018,Cheng2019,cheng2019regression,husch2020improved} Gaussian process (GP) \cite{GPbook} describes a  prior distribution of random functions such that, for any finite number of possible inputs $\boldsymbol{x}=\{x_1,\dots,x_n\}$, the function values $f(\boldsymbol{x})$ has a multivariate Gaussian distribution
\begin{equation}
    f(\boldsymbol{x}) \sim N(0, K(\boldsymbol{x}, \boldsymbol{x})),
\end{equation}
where $K(\boldsymbol{x}, \boldsymbol{x})$ is the kernel matrix.
Assuming the training data $(\mathbf{X}, \mathbf{y})$ has a Gaussian distributed noise with variance $\sigma_n^2$ (also referred as Gaussian likelihood), i.e., $\mathbf{y}\sim N(f,\Sigma_n)$, where $\Sigma_n=\sigma^2_n I$,
the GPR prediction $f(\mathbf{X}^\star)$ for the test points is a multivariate Gaussian distribution with the mean (prediction) and variance (uncertainty) as 
\begin{equation} \label{eq:GPR_pred}
\begin{aligned} 
    \mathbb{E}[f(\mathbf{X}^\star)] &= K(\mathbf{X}^\star,\mathbf{X})(K(\mathbf{X}, \mathbf{X})+\Sigma_n)^{-1} \mathbf{y}
    \\
    \text{Var}[f(\mathbf{X}^\star)] &= K(\mathbf{X}^\star,\mathbf{X})(K(\mathbf{X}, \mathbf{X})+\Sigma_n )^{-1}K(\mathbf{X},\mathbf{X}^\star).
\end{aligned}
\end{equation}

Although GPR only requires a few hundred molecules to achieve highly-accurate models, it is difficult to increase the training size due to the high computational complexity ($O(N^3)$) of full GPR. \cite{Cheng2019,husch2020improved}
To scale up the MOB-ML training, a regression-with-clustering framework is introduced by constructing a local regressor within each grouping of points. 
The resulting clusters from GMM agree with the chemically intuitive groupings of MO types \cite{cheng2022}, and thus could be directly applied to facilitate the construction of local regression models of pair dipole moments.
In addition, a scalable exact GPR regressor, termed Alternative Black-box Matrix-Matrix Multiplication (AltBBMM) \cite{sun2021molecular}, has been developed to lower the computational cost of each local regression.
Ref. \citenum{cheng2022} accesses different combinations of clustering methods and local regressors and concludes the most efficient protocol is local regression by AltBBMM with GMM (GMM/GPR), which achieves the state-of-the-art accuracy on drug-like benchmark organic datasets.

\subsection{GPR with derivatives for ML response and rotational equivariance}
\label{sec:derivative}

Several previous studies have recognized the importance of rotational equvariance for ML framework to efficiently learn molecular dipole moments. \cite{painn,orbnetequi}
For any rotational operator $\hat{U}$, a function $\boldsymbol{g}:\boldsymbol{x} \rightarrow \boldsymbol{g}(\boldsymbol{x})$ is rotationally equivariant if
\begin{equation}
\label{eq:equivariant}
    \boldsymbol{g}(\hat{U} \boldsymbol{x}) = \hat{U} \boldsymbol{g}(\boldsymbol{x}).
\end{equation}
From a physics perspective, the rotational equivariance guarantees that the predicted property will rotate correspondingly with the rotation of the system. 
Therefore, the rotational equivariance property is required for any tensorial molecular properties, such as force, dipole moment, and polarizability. 
In addition, the molecular energy is rotationally invariant, i.e., remains constant with the rotation of the system. 
The conditions for energy and dipole models can be formulated as following:
\begin{equation} \label{eq:MOBML_rotation}
\begin{cases}
    \epsilon^{\text{ML}}[\boldsymbol{f}^\epsilon_{ij}] = \epsilon^{\text{ML}}[\hat{U}\boldsymbol{f}^\epsilon_{ij}]
    \\
    \boldsymbol{\mu}^{\text{ML}}[\hat{U}\boldsymbol{f^\mu}_{ij}] = \hat{U}\boldsymbol{\mu}^{\text{ML}}[\boldsymbol{f^\mu}_{ij}].
\end{cases}
\end{equation}

For the MOB features, when applying a rotation operator $
\hat{U}$, the energy and dipole feature sets satisfy the relationship  $\hat{U}\boldsymbol{f}_{ij}^\epsilon=\boldsymbol{f}_{ij}^\epsilon$ and
$\hat{U}\boldsymbol{f^\mu}_{ij} = \{\boldsymbol{f}^\epsilon_{ij}, U\nabla_{\pmb{\mathcal{E}}} \boldsymbol{f}^\epsilon_{ij} \}$, respectively, where $U$ is the matrix representation of $\hat{U}$. Therefore, the MOB-ML energy model is always rotationally invariant for any regressor. However, it remains challenging and requires a special ML algorithm design to make the MOB-ML dipole model rotationally equivariant for a greater learning efficiency.

Assuming the training energy and dipole sets are $(\mathbf{X_\epsilon}, \mathbf{y_\epsilon}) = \{\boldsymbol{f}^{\boldsymbol{\epsilon}}_{ij}, \epsilon_{ij}\}$ and $(\mathbf{X_\mu}, \mathbf{y_\mu})=\{\boldsymbol{f}^{\boldsymbol{\mu}}_{ij}, \mathbf{\mu}_{ij}\}$, respectively, we introduce a rotationally equivariant GPR with derivatives formalism that could accurately learn dipole moment and energy separately or simultaneously.

A single-task energy model could be directly learnt using the naive GPR in Eq.~\ref{eq:GPR_pred} with the prior distribution
\begin{equation}
    \mathbf{\epsilon}^{\text{ML}}(\mathbf{X}_\epsilon)
    \sim 
    N(0, K_\mathbf{\epsilon}(\mathbf{X}_\epsilon, \mathbf{X}_\epsilon)).
\end{equation}
Since the derivative of a GP is also a GP \cite{GPbook,parzen1999stochastic}, dipole moments can be regressed by GPR with the prior distribution of $\nabla_{\pmb{\mathcal{E}}} \mathbf{\epsilon}^{\text{ML}}(\mathbf{X}_\epsilon)$:
\begin{equation}
    \boldsymbol{\mu}^{\text{ML}}(\mathbf{X_\mu}) = \nabla_{\pmb{\mathcal{E}}} \mathbf{\epsilon}^{\text{ML}}(\mathbf{X}_\epsilon)
    \sim 
    N(0, K_\mathbf{\mu}(\mathbf{X_\mu}, \mathbf{X_\mu})).
\end{equation}
$\mu^{\text{ML}}(\boldsymbol{x})$ and the corresponding kernel matrix $K_\mathbf{\mu}$ could be written as following:
\begin{align}
    \boldsymbol{\mu}^{\text{ML}}(\mathbf{X_\mu}) & = \nabla_{\pmb{\mathcal{E}}} \mathbf{X}_\epsilon \cdot \nabla_{\mathbf{X}_\epsilon} \epsilon^{\text{ML}} (\mathbf{X}_\epsilon), \nonumber \\ 
    K_\mathbf{\mu}(\mathbf{X_\mu}, \mathbf{X_\mu}) &= \nabla_{\pmb{\mathcal{E}}} \mathbf{X}_\epsilon \nabla_{\pmb{\mathcal{E}}} \mathbf{X_\epsilon}
    \cdot K_\mathbf{\epsilon}^{(1,2)}(\mathbf{X_\epsilon}, \mathbf{X_\epsilon}),
\end{align}
where the superscripts of $K_\mathbf{\epsilon}$ represent the derivatives to the arguments, e.g. $K^{(1,2)}(x_1, x_2) = \nabla_{x_1} \nabla_{x_2} K(x_1, x_2)$, 
Since $\nabla_{\pmb{\mathcal{E}}} \mathbf{X}_\epsilon$ will produce the derivative terms $\{\nabla_{\pmb{\mathcal{E}}} \boldsymbol{f}^\epsilon_{ij}\}$, including these derivatives in the dipole feature set is necessary to model  $\mathbf{\mu}^{\text{ML}}$. This mathematical deduction agrees with the physical intuition discussed in Sec.~\ref{sec:dipole_feature}.

By using the Gaussian likelihood $\mathbf{y_\mu} \sim N(\boldsymbol{\mu}^{\text{ML}},\Sigma_\mu)$ with $\Sigma_\mu = \sigma_\mu^2 I$, for a set of test points $\mathbf{X^\star_\mu}$, the prediction mean and variance of the GPR with derivatives can be evaluated using Eq.~\ref{eq:GPR_pred} as
\begin{equation} \label{eq:GPRd_pred}
\begin{aligned} 
    \mathbb{E}[\boldsymbol{\mu}^{\text{ML}}(\mathbf{X^\star_\mu})] =& K_\mu(\mathbf{X^\star_\mu},\mathbf{X_\mu})(K_\mu(\mathbf{X_\mu}, \mathbf{X_\mu})+\Sigma_\mu)^{-1} \mathbf{y_\mu}
    \\
    \text{Var}[\boldsymbol{\mu}^{\text{ML}}(\mathbf{X^\star_\mu})] =& K_\mu(\mathbf{X^\star_\mu},\mathbf{X_\mu})(K_\mu(\mathbf{X_\mu}, \mathbf{X_\mu})+\Sigma_\mu )^{-1}
    \\
    &K_\mu(\mathbf{X_\mu},\mathbf{X^\star_\mu}),
\end{aligned}
\end{equation}
where
\begin{equation}
    K_\mu(\mathbf{X_\mu},\mathbf{X^\star_\mu})= K_\mu(\mathbf{X^\star_\mu},\mathbf{X_\mu})^\text{T}=\nabla_{\pmb{\mathcal{E}}} \mathbf{X}_\epsilon \nabla_{\pmb{\mathcal{E}}} \mathbf{X^\star_\epsilon}
\cdot K_\mathbf{\epsilon}^{(1,2)}(\mathbf{X_\epsilon}, \mathbf{X^\star_\epsilon}).
\end{equation}

GPR with derivatives can be generalized to the multi-task learning of $\epsilon^{\text{ML}}$ and $\boldsymbol{\mu}^{\text{ML}}$ simultaneously.
In such case, their joint distribution is also a GP with the predictive mean and variance as
\begin{equation} \label{GPR:pred_m}
\begin{aligned}
    \mathbb{E}
    \left[
    \begin{array}{c}
    \epsilon^{\text{ML}}(\mathbf{X^\star_\mu})
    \\
    \boldsymbol{\mu}^{\text{ML}}(\mathbf{X^\star_\mu})
    \end{array}
    \right] =& K_{\epsilon\mathbf{\mu}}(\mathbf{X^\star_\mu},\mathbf{X_\mu})(K_{\epsilon\mathbf{\mu}}(\mathbf{X_\mu}, \mathbf{X_\mu})+\Sigma_{\epsilon\mathbf{\mu}})^{-1} \mathbf{y_{\epsilon\mathbf{\mu}}}
    \\
    \text{Var}
    \left[
    \begin{array}{c}
    \epsilon^{\text{ML}}(\mathbf{X^\star_\mu})
    \\
    \boldsymbol{\mu}^{\text{ML}}(\mathbf{X^\star_\mu})
    \end{array}
    \right] =& K_{\epsilon\mathbf{\mu}}(\mathbf{X^\star_\mu},\mathbf{X_\mu})(K_{\epsilon\mathbf{\mu}}(\mathbf{X_\mu}, \mathbf{X_\mu})+\Sigma_{\epsilon\mathbf{\mu}})^{-1}
    \\
    &K_{\epsilon\mathbf{\mu}}(\mathbf{X_\mu},\mathbf{X^\star_\mu}),
\end{aligned}
\end{equation}
where
\begin{equation} \label{eq:kernel_m}
\begin{aligned}
    &\mathbf{y}_{\epsilon\mathbf{\mu}} = \left[\begin{array}{c} \mathbf{y}_{\epsilon} \\ \mathbf{y}_{\mathbf{\mu}} \end{array} \right],\ 
    \Sigma_{\epsilon\mathbf{\mu}} = \left[\begin{array}{cc} \sigma_\epsilon^2 I & 0 \\ 0 & \sigma_\mathbf{\mu}^2 I \end{array} \right],\\
    &
    K_{\epsilon\mathbf{\mu}}(\mathbf{X_\mu}, \mathbf{X_\mu^\star}) = K_{\epsilon\mathbf{\mu}}(\mathbf{X_\mu^\star}, \mathbf{X_\mu})^\text{T} =
    \\
    &
    \left[
    \begin{array}{cc}
    K_\epsilon(\mathbf{X_\epsilon}, \mathbf{X_\epsilon^\star}) & \nabla_{\pmb{\mathcal{E}}} \mathbf{X_\epsilon^\star} \cdot K^{(2)}_\epsilon(\mathbf{X_\epsilon}, \mathbf{X_\epsilon^\star})
    \\
    \nabla_{\pmb{\mathcal{E}}} \mathbf{X_\epsilon} \cdot K^{(1)}_\epsilon(\mathbf{X_\epsilon}, \mathbf{X_\epsilon^\star}) & \nabla_{\pmb{\mathcal{E}}} \mathbf{X_\epsilon} \nabla_{\pmb{\mathcal{E}}} \mathbf{X_\epsilon^\star} \cdot K^{(1,2)}_\epsilon(\mathbf{X_\epsilon}, \mathbf{X_\epsilon^\star})
    \end{array}
    \right]
\end{aligned}
\end{equation}
and $K_{\epsilon\mathbf{\mu}}(\mathbf{X_\mu}, \mathbf{X_\mu})$ could be evaluated by replacing $\mathbf{X_\mu^\star}$ to $\mathbf{X_\mu}$.

Although there might be other rotationally equivariant GPR frameworks that provide models with similar accuracy, they might not ensure the learnt dipole model is a derivative of the energy model.
As an analogy to the response in physics, it is desirable for the ML model of $h$ and its response property model $\boldsymbol{g}$ to satisfy Eq.~\ref{eq:ml_response}, termed as "ML response relationship".
\begin{equation}
\label{eq:ml_response}
    \boldsymbol{g}(\boldsymbol{x}) = \nabla h(\boldsymbol{x}).
\end{equation}
This relationship requires model $\boldsymbol{g}$ to be conservative (or curl-free), i.e., $\nabla \times \boldsymbol{g} = \boldsymbol{0}$ \cite{marsden2003vector,GDML}. 
In this study, we apply this physically driven GPR with derivatives formalism to satisfy this response relationship.
The detailed proofs to show that both single-task and multi-task GPR with derivatives satisfy the rotational equivariance (Eq.~\ref{eq:MOBML_rotation}). 
Without any specification, we adapt the single-task framework of GPR with derivatives for all the following training. The performance comparisons of single-task and multi-task models are demonstrated in Sec.~\ref{sec:water_small}. 

\section{Computational details}
\subsection{Data generation}
For water and fourteen small molecules, we sample 200 total configurations at 50 fs intervals from \textit{ab initio} molecular dynamics (AIMD) trajectories using the \textsc{entos qcore} \cite{manby2019entos} software. The structures of the QM9 dataset and the four different series of peptides are directly obtained from Ref.~\citenum{ramakrishnan2014quantum} and  Ref.~\citenum{veit2020predicting}, respectively.
All the HF calculations and MOB-ML features generations are performed by the \textsc{entos qcore} software.
The HF calculations are using the cc-pVTZ basis \cite{Dunning1989} and the cc-pVTZ-JKFIT density fitting basis \cite{weigend_fully_2002}.
The valence-occupied and valence-virtual MOs are then localized using the Boys-Foster localization scheme. \cite{Boys1960,husch2020improved}
The derivatives of the orbitals with respect to electric fields are calculated through the coupled-perturbed HF calculations implemented in the \textsc{entos qcore} software.
According to the chain rule, the feature derivatives with respect to electric fields are then calculated from the orbitals and orbital derivatives.

The HF orbitals are then imported into the Molpro 2018.0 \cite{MOLPRO} package via the \texttt{matrop} functionality to generate the local MP2 \cite{LMP2} pair energies.
The frozen core approximation is used in the local MP2 calculations.
The derivatives of the pair energies with respect to electric fields are implemented and calculated in the Molpro package following Ref. \onlinecite{LMP2gradient}.

\subsection{Machine learning protocols}
For all the datasets, we separately learn the energies and dipole moments using the energy feature set and dipole feature set, respectively. For the water and the small molecule datasets, the results for multi-task models, i.e., dipole + energy models, that learn both tasks simultaneously are also included for comparison.
Table \ref{tab:feature_usage} summarizes the usage of energy and dipole features in this work. We unsupervisedly cluster the MOs represented by energy features instead of separately clustering energy and dipole feature space, i.e., the clustering models are identical for the energy and dipole learning with the same training sets. 
Feature selection is performed before all the models on energy labels using the random forest regression implementation in the \textsc{Scikit-learn} \cite{scikit-learn} package following the protocol in Ref.~\citenum{Cheng2019}. For the dipole learning, we use the selected energy features and their derivatives as the selected dipole features. 

\begin{table}[bhtp]
\caption{Usages of different feature sets in different learning models.}
\begin{ruledtabular}
\begin{tabular}{c|c|cc}
ML model & Feature set & Learning task \\\hline
Clustering (GMM) & Energy features & All energy \& dipole models \\ \hline
\multirow{3}{*}{Regression (GPR)} & Energy features & Energy (single-task)\\ \cline{2-3}
& \multirow{2}{*}{Dipole features} & Dipole (single-task)\\
& & Dipole + energy (multi-task)
\end{tabular}   
\end{ruledtabular}
\label{tab:feature_usage}
\end{table}

We apply the AltBBMM algorithm as the default GP regressor, and reimplement the GMM with full covariance matrix following \textsc{Scikit-learn} using \textsc{CuPy} \cite{cupy_learningsys2017} to enable multi-GPU training. The implementations for both algorithms are available online at \url{https://github.com/SUSYUSTC/BBMM.git}
For all GPR and GMM/GPR, we employ the Mat\'ern 5/2 kernel with white noise regularization \cite{GPbook}. The parameters used in training for GPR and GMM/GPR are further discussed in the Supporting information. 

All the results for water and small molecule datasets are collected with GPR without clustering. The local GPR with GMM unsupervised clustering is applied to scale the MOB-ML training in the QM9 dataset. In this work, we follow the same clustering protocol introduced in Ref.~\citenum{cheng2022} to generate the GMM models. 
The GMM model is initialized by K-means clustering, and its number of clusters is automatically determined by minimizing the Bayesian information criterion.
GMM could not be performed to cluster 50000 and 110000 QM9 molecules due to limited memory, and we thus apply the GMM model trained on 20000 QM9 molecules to approximate the clustering results of these two models. To reduce the learning costs, we also apply the same capping strategy described in Ref. \citenum{cheng2019regression}. For the clusters containing a large number of points, we randomly select training points with the capping size defined ahead to regress these local GPR. 
The capping size is 1000000 and 300000 pairs for dipole and  energy learning, respectively.

\section{Results and Discussions}
\subsection{Dipole moment learning for small molecules via MOB-ML}
\label{sec:water_small}
\begin{figure}[hbtp]
    \centering
    \includegraphics[width=0.9\columnwidth]{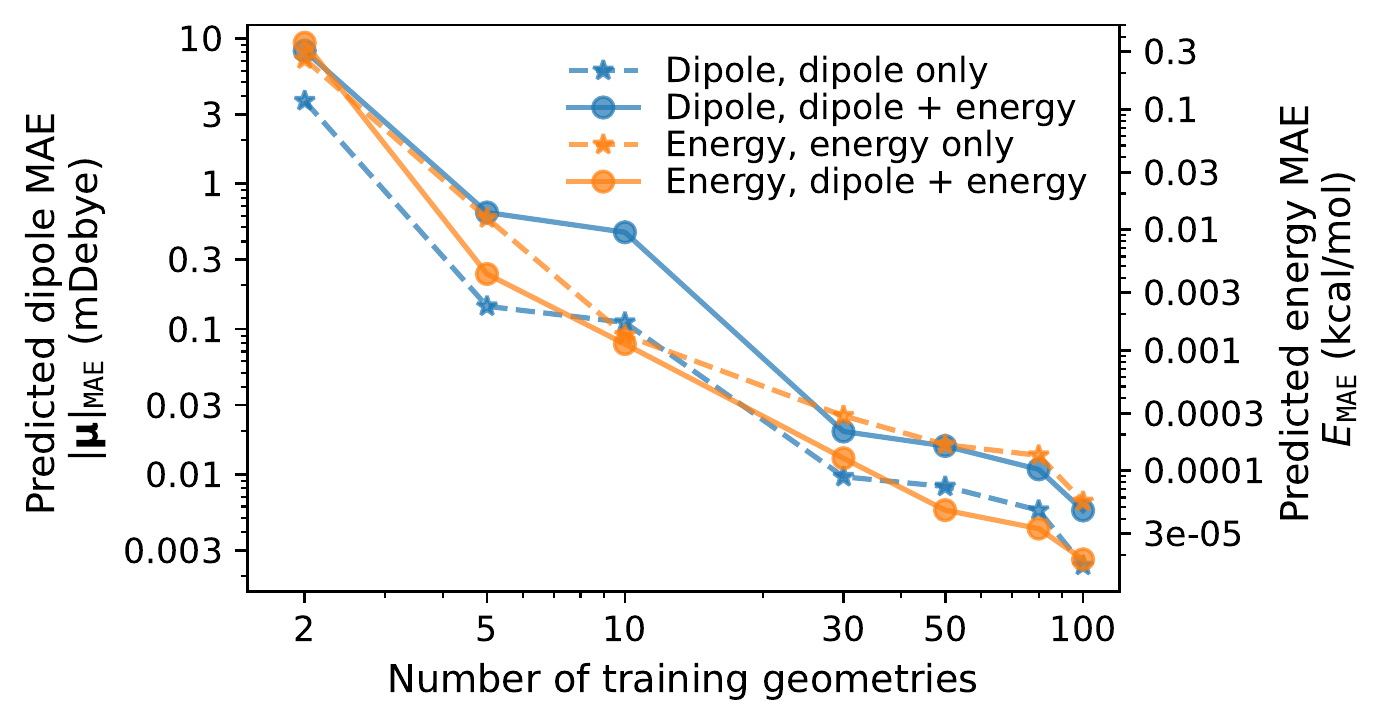}
    \caption{Prediction accuracies of dipole moments and energies of water using a reference theory of MP2/cc-pVTZ. The predicted MAEs are plotted versus the number of training geometries on a log-log scale ("learning curves"). The MOB-ML models for dipole moments and energies are constructed by training on the labels individually (dipole model and energy model) or simultaneously using multi-task learning (dipole $+$ energy model). The primary and secondary y-axises represent the prediction MAE of dipole moments $|\boldsymbol{\mu}|_{\text{err}}$ in milli-Debye (mDebye) and the prediction MAE of energy in kcal/mol, separately. The results are also summarized in Table S1 in the Supporting Information (SI)}    
    \label{fig:water_lc}
\end{figure}

To demonstrate the ability to learn dipole moments using the MOB representations, we first test the prediction accuracies of MOB-ML on water and other small molecules. Figure \ref{fig:water_lc} displays the mean absolute errors (MAEs) of dipole moments ($|\boldsymbol{\mu}|$) and molecular energies $E$ for water learnt by single-task and multi-task models. The sizes of the training set are varied from 2 to 100 geometries, and the test set is composed of 100 geometries not included in any training sets. MAEs at the same order of reference data accuracy for molecular energies are achieved by training on 100 geometries using energy labels ($5.54 \times 10^{-5}$ kcal/mol) and dipole combined with energy labels ($1.84 \times 10^{-5}$ kcal/mol). Across all the training sizes, single-task and multi-task models provide similar accuracies for molecular energies, but dipole models provide much better dipole predictions than the dipole + energy model.

\begin{table*}[bhtp]
\caption{Predicted error of the MOB-ML training on dipole only or dipole and energy together on different small molecules. All the models are trained on 50 configurations and tested on the rest 50 configurations using AltBBMM as the regressor.}
\begin{ruledtabular}
\begin{tabular}{m{1.3cm}|ll|ll|llll}
\multirow{3}{*}{System} & \multicolumn{2}{c|}{Dipole only}   & \multicolumn{2}{c|}{Energy only} & \multicolumn{4}{c}{Dipole + Energy}  \\
  & $|\boldsymbol{\mu}|_{\text{MAE}}$ & $|\boldsymbol{\mu}|_{\text{Max}}$ & $E_{\text{MAE}}$ & $E_{\text{Max}}$ & $|\boldsymbol{\mu}|_{\text{MAE}}$ & $|\boldsymbol{\mu}|_{\text{Max}}$ & $E_{\text{MAE}}$ & $E_{\text{Max}}$ \\
& (mDebye) & (mDebye) & (kcal/mol)  & (kcal/mol)    & (mDebye)  & (mDebye) & (kcal/mol)       & (kcal/mol)    \\ \hline
\ch{CH4} & 0.024 & 0.088 & 0.0005 & 0.002 & 0.040 & 0.183 & 0.003 & 0.015 \\
\ch{NH3} & 0.030 & 0.109 & 0.0007 & 0.003 & 0.080 & 0.362 & 0.002 & 0.008 \\
\ch{HF} & 0.00008 & 0.0004 & 0.00005 & 0.00009 & 0.0001 & 0.0005 & 0.00007 & 0.0001 \\
\ch{CO} & 0.004 & 0.016 & 0.00009 & 0.0006 & 0.010 & 0.031 & 0.001 & 0.003 \\
\ch{CH2O} & 0.030 & 0.213 & 0.0005 & 0.003 & 0.049 & 0.144 & 0.003 & 0.010 \\
\ch{HCN} & 0.052 & 0.643 & 0.0002 & 0.002 & 0.137 & 4.898 & 0.005 & 0.016 \\
\ch{C2H4} & 0.151 & 1.112 & 0.004 & 0.012 & 0.260 & 3.555 & 0.011 & 0.038 \\
\ch{C2H6} & 0.339 & 0.941 & 0.014 & 0.069 & 0.482 & 4.069 & 0.028 & 0.283 \\
\ch{CH3OH} & 1.143 & 12.281 & 0.018 & 0.084 & 1.430 & 17.310 & 0.020 & 0.101 \\
\ch{CH2F2} & 2.180 & 16.847 & 0.080 & 0.884 & 3.529 & 24.410 & 0.060 & 0.270 \\
\ch{C3H8} & 0.565 & 3.731 & 0.035 & 0.135 & 0.998 & 4.682 & 0.047 & 0.331 \\
n-Butane & 0.912 & 3.338 & 0.026 & 0.089 & 1.842 & 6.818 & 0.076 & 0.301 \\
Isobutane & 0.812 & 2.588 & 0.047 & 0.196 & 1.883 & 5.588 & 0.071 & 0.312 \\
\ch{C6H6} & 2.433 & 9.277 & 0.039 & 0.113 & 3.403 & 20.020 & 0.053 & 0.129
  \end{tabular}   
   \end{ruledtabular}
    \label{tab:small_mols}
\end{table*}
Table \ref{tab:small_mols} lists the MAEs and maximum errors (Max) of dipole moments and total energies from single-task and multi-task models training on 50 geometries and testing on different 50 geometries for small molecules with different sizes. MOB-ML provides very accurate predictions for all the test molecules. Comparing the results of molecules with different molecular sizes sharing similar MO properties, such as \ch{CH4}, \ch{C2H6}, and \ch{C3H8}, it is clear that the larger molecules have much bigger errors than the smaller ones for both dipole and energy owing to the increasing number of pairwise contributions to the final result. 
The total errors should scale linearly with the increase of molecular size by summing up predicted pairwise contribution with Gaussian distributed pairwise errors. For the systems that share similar numbers of MOs, such as \ch{C2H4} and \ch{C2H6}, the more rigid molecule (\ch{C2H4}) is easier to learn for both dipole and energy.

Learning dipole moments and molecular energies simultaneously does not always provide better prediction accuracies of both dipole and energy for most of the molecules (12 out of total 14 molecules) and need higher computational costs since it trains more points within a model. 
This observation indicates that dipole vectors and energies of each pair of MOs might vary independently as functions of MOB features, and therefore no mutual supervision could be provided by multi-task learning.

\begin{figure}
    \centering
    \includegraphics[width=0.8\columnwidth]{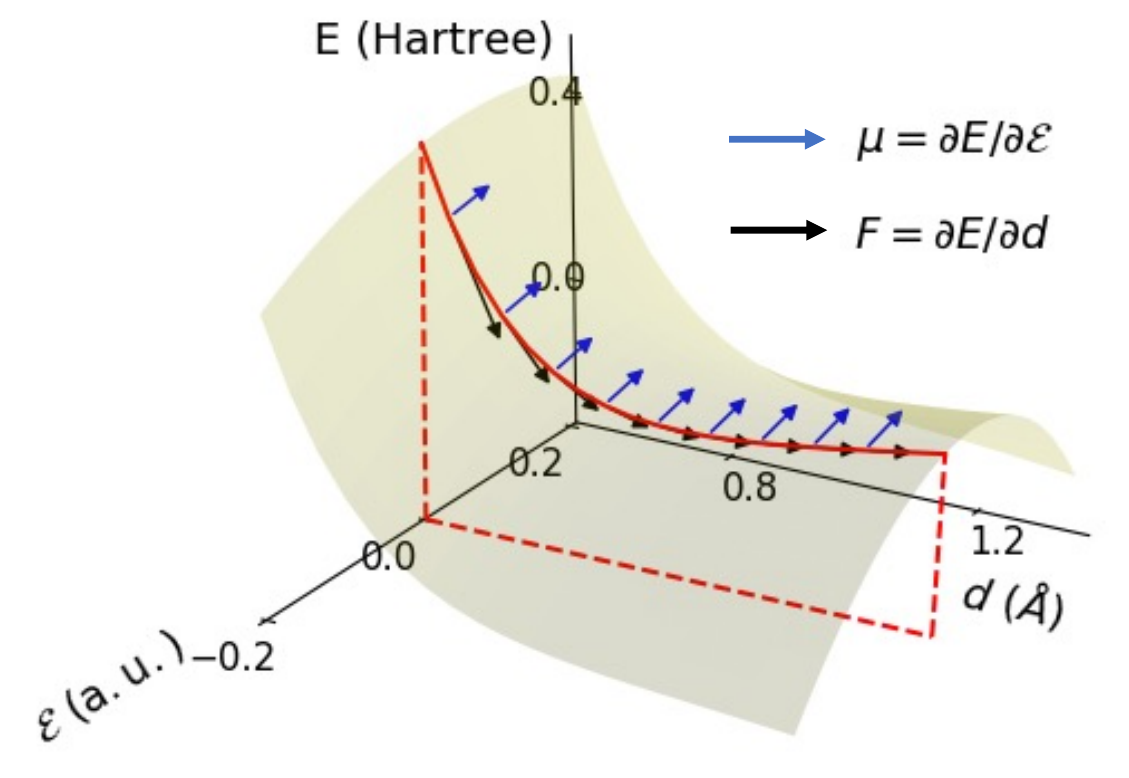}
    \caption{Relative MP2/cc-pVTZ total energy of a water molecule as a function of one of its O-H bond lengths $d$ and the strength of applied electric field $\mathcal{E}$ along the bond direction. The bond angle and the other O-H bond length are fixed to the equilibrium value. The relative energy is shifted to 0 at equilibrium geometry with $\mathcal{E}=0$. The red curve corresponds to the energy surface at $\mathcal{E}=0$. The dashed red lines represent the projection of the red curve to $d$-$\mathcal{E}$ plane. The blue and black arrows illustrate the direction of the derivatives to get dipole moments and forces.}
    \label{fig:water_3d}
\end{figure}

From a theoretical perspective, the ML response relationship could improve the learning efficiency of forces, but is not expected to enhance the learning efficiency of the dipole model with zero electric field. Therefore, the dipole + energy model for these small molecules cannot provide better accuracies than only training on dipoles. Furthermore, the inclusion of non-correlated tasks in the multi-task models further complicates the learning process and even leads to deterioration in training. 
Figure ~\ref{fig:water_3d} could help explain this observation by using the water molecule as an example. The relative water MP2/cc-pVTZ total energy is plotted as a function of one of the O-H bond lengths and the strength of the external electric field along the bond direction. We fix the bond angel and only change one of the O-H bond lengths (red curve) with $\mathcal{E}=0$ to facilitate the understanding. The training set can be treated as samples of this simplified potential. We note that all the bond lengths and angles vary in the actual water dataset. 
From these data, the energy information can then directly infer the force information (black arrows) since the derivative with respect to bond length $d$ could be estimated as differences between two sampled training points. However, since there is no change along the axis of the electric field $\pmb{\mathcal{E}}$, no estimated information of dipole moments (blue arrows) is available from the training set.

\subsection{MOB-ML for dipole moments and energies of organic molecules in QM9}

\begin{figure*}[htbp]
    \centering
    \includegraphics[width=2\columnwidth]{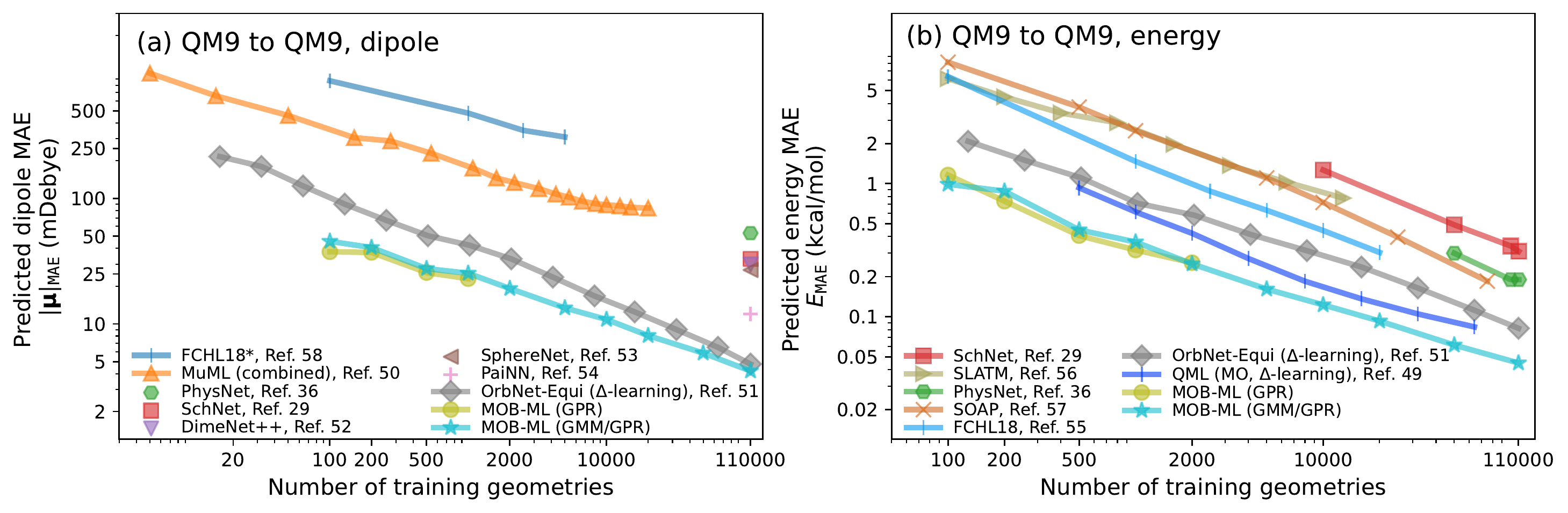}
    \caption{Prediction accuracies of dipole moments and energies of test QM9 molecules trained on QM9 molecules using MOB-ML. The single-task MOB-ML models for (a) dipole moments and (b) energies are constructed by training on the labels individually. For the methods that only report the prediction errors of the models training on 110,000 molecules, we plot their results as scatters with different shapes. The learning curves of other literature methods trained on QM9 properties computed using B3LYP/6-31G(2df,p) level of theory \cite{ramakrishnan2014quantum} are also plotted for comparison. The prediction MAEs of dipole moments $|\boldsymbol{\mu}|_{\text{err}}$ and energy are in milli-Debye (mDebye) and kcal/mol units, respectively. All the results of MOB-ML are also listed in Table S2 in SI.}    
    \label{fig:qm9_to_qm9}
\end{figure*}
In our previous studies, we have illustrated the excellent accuracy and transferability of MOB-ML to learn molecular energies using two thermalized organic molecule datasets, i.e., QM7b-T and GDB-13-T \cite{Cheng2019,cheng2019regression,husch2020improved,cheng2022}. In this study, we systematically examine the learning performance of MOB-ML for both the dipole and energy using the benchmark organic chemistry dataset QM9\cite{ramakrishnan2014quantum}, which contains optimized structures of 133885 molecules with up to nine heavy atoms (HAs) of C, O, N, and F. QM9 is a standard benchmark dataset that has been assessed in many different literature studies. \cite{OQML,veit2020predicting,unke2019physnet,schutt2017schnet,klicpera2020fast,liu2022spherical,painn,orbnetequi,qiao2020orbnet,faber2018alchemical,huang2020quantum,Karandashev2022,bartok2017machine}  Figure~\ref{fig:qm9_to_qm9} displays the predicted MAEs for dipole moments (in mDebye) and energies (in kcal/mol) as functions of number of training geometries on a log-log scale (learning curves). Since our GPR regression, AltBBMM, can only train at most 1 million points, we collect the results of MOB-ML (GPR) up to training on 1000 and 2000 dipole moments and molecular energies, respectively. The application of GMM/GPR scales the training of dipole moments and energy to the same training size (at most 110,000 QM9 molecules) as the literature models. The test sets of MOB-ML approaches remain the same across the entire learning curve with a size of 11,843 molecules. Different literature approaches computed at the B3LYP/6-31G(2df,p) level of theory are included for comparison.

For dipole moments in Fig.~\ref{fig:qm9_to_qm9}a, we compare the results from MOB-ML with those from the state-of-the-art literature (SOTA) methods, including FCHL18* \cite{OQML}, MuML (combined) \cite{veit2020predicting}, PhysNet \cite{unke2019physnet}, SchNet \cite{schutt2017schnet}, DimeNet++ \cite{klicpera2020fast}, SphereNet \cite{liu2022spherical}, PaiNN \cite{painn}, and OrbNet-Equi \cite{orbnetequi}. 
It is clear that both MOB-ML regressed by AltBBMM (MOB-ML(GPR)) and MOB-ML regressed by GMM clustering with local AltBBMM (MOB-ML(GMM/GPR)) outcompete other literature methods in the low-data learning regime (training set smaller than 2000 molecules). MOB-ML/GPR and MOB-ML(GMM/GPR) achieve similar MAEs of 37.61 and 45.58 mDebye, respectively, by training on only 100 molecules. This indicates that the introduction of unsupervised clustering does not affect the accuracy of MOB-ML in learning dipole moments.
Meanwhile, the second-best OrbNet-Equi learnt by $\Delta$-learning (OrbNet-Equi ($\Delta$-learning)) \cite{orbnetequi} requires 1024 molecules to reach the same level of accuracy. 
However, OrbNet-Equi ($\Delta$-learning) models are improved faster with an increasing number of data points with the deepest slope of the learning curve relative to other methods. 
When there are enough examples in the training set (110000 training molecules), OrbNet-Equi ($\Delta$-learning) could reach a slightly worse MAE of 4.78 mDebye than MOB-ML(GMM/GPR) (4.21 mDebye).

Similarly, the prediction errors of energies from MOB-ML approaches are compared with SchNet \cite{schutt2017schnet}, SLATM \cite{huang2020quantum}, PhysNet \cite{unke2019physnet}, SOAP \cite{bartok2017machine}, FCHL18 \cite{faber2018alchemical},  OrbNet-Equi \cite{orbnetequi}, and QML \cite{Karandashev2022} in Fig.~\ref{fig:qm9_to_qm9}b. MOB-ML (GPR) and MOB-ML (GMM/GPR) are still provide the best sets of results across all the training sizes. MOB-ML (GMM/GPR) achieves accuracies of 0.99 kcal/mol and 0.045 kcal/mol with only 100 and 110000 training molecules, respectively. Both numbers are the current best in this field. QML with an orbital based features and ($\Delta$-learning) (QML (MO, $\Delta$-learning)) and OrbNet-Equi ($\Delta$-learning) are other two most accurate approaches. 

The top approaches to predict dipole moments and energies, i.e., MOB-ML, OrbNet-Equi ($\Delta$-learning), and QML (MO, $\Delta$-learning), are further compared here from a  theoretical perspective. 
To achieve the best accuracy, all the three approaches apply the idea of "$\Delta$-learning" by predicting the differences between low-level and high-level theories instead of directly predicting. 
In addition, all three approaches are orbital-based ML approaches that adapt features related to energy matrix elements (FJK features), and these features are considered to contain high-quality quantum-level information to make the ML map easier. 
Although the three approaches share several similarities, their differences might explain their prediction accuracy differences.
Firstly, MOB-ML predicts the results from wavefunction theories using HF computations with the same basis, while OrbNet-Equi and QML (MO) predict the results from DFT using GFN-xTB and minimal basis HF computations, respectively. 
Since HF computed with cc-pVTZ basis set is more expensive and contains more accurate information about the orbitals than GFN-xTB and minimal basis HF, MOB-ML is more expensive in evaluation than the other two methods. 
MOB-ML explicitly decomposes the differences between low- and high-level theory results onto MOs and learns these pairwise contributions using GPR, while OrbNet-Equi and QML (MO) directly learn these differences by implicitly decomposing them to each kernel in kernel ridge regression (KRR) or nodes in graph neural network (GNN) by carefully designing the ML frameworks. 
This explicit decomposition brings an accuracy gain to MOB-ML but limits the applications of MOB-ML to decomposable properties. On the other hand, QML (MO) and OrbNet-Equi are able to predict more different molecular properties. 

\begin{figure}[htbp]
    \centering
    \includegraphics[width=0.95\columnwidth]{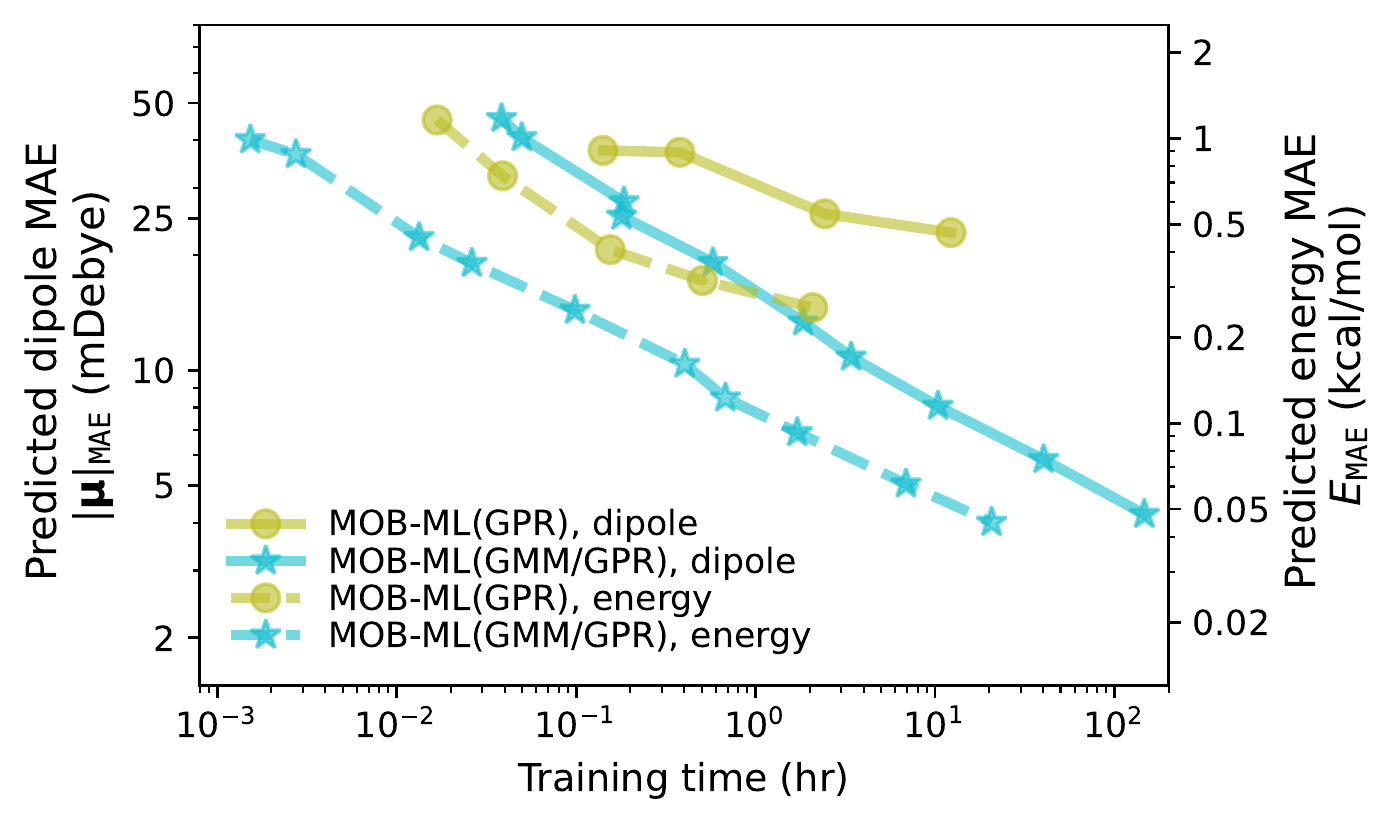}
    \caption{Learning costs of MOB-ML approaches for QM9 models shown in Fig.~\ref{fig:qm9_to_qm9}.
    The prediction MAEs for dipole moments and energies of QM9 test molecules are plotted as functions training hours using 8 GPUs on log-log scales. The primary axis (left axis) labels the MAEs of the dipole moments, and the secondary axis (right axis) shows the MAEs of energies. Results for dipole and energy models are plotted in solid and dashed lines, respectively. Different colors represent different learning protocols and match the ones in Fig.~\ref{fig:qm9_to_qm9}}     
    \label{fig:timing}
\end{figure}

\subsection{Timing and learning efficiency of GMM/AltBBMM with derivatives}

\begin{figure*}[htbp]
    \centering
    \includegraphics[width=1.5\columnwidth]{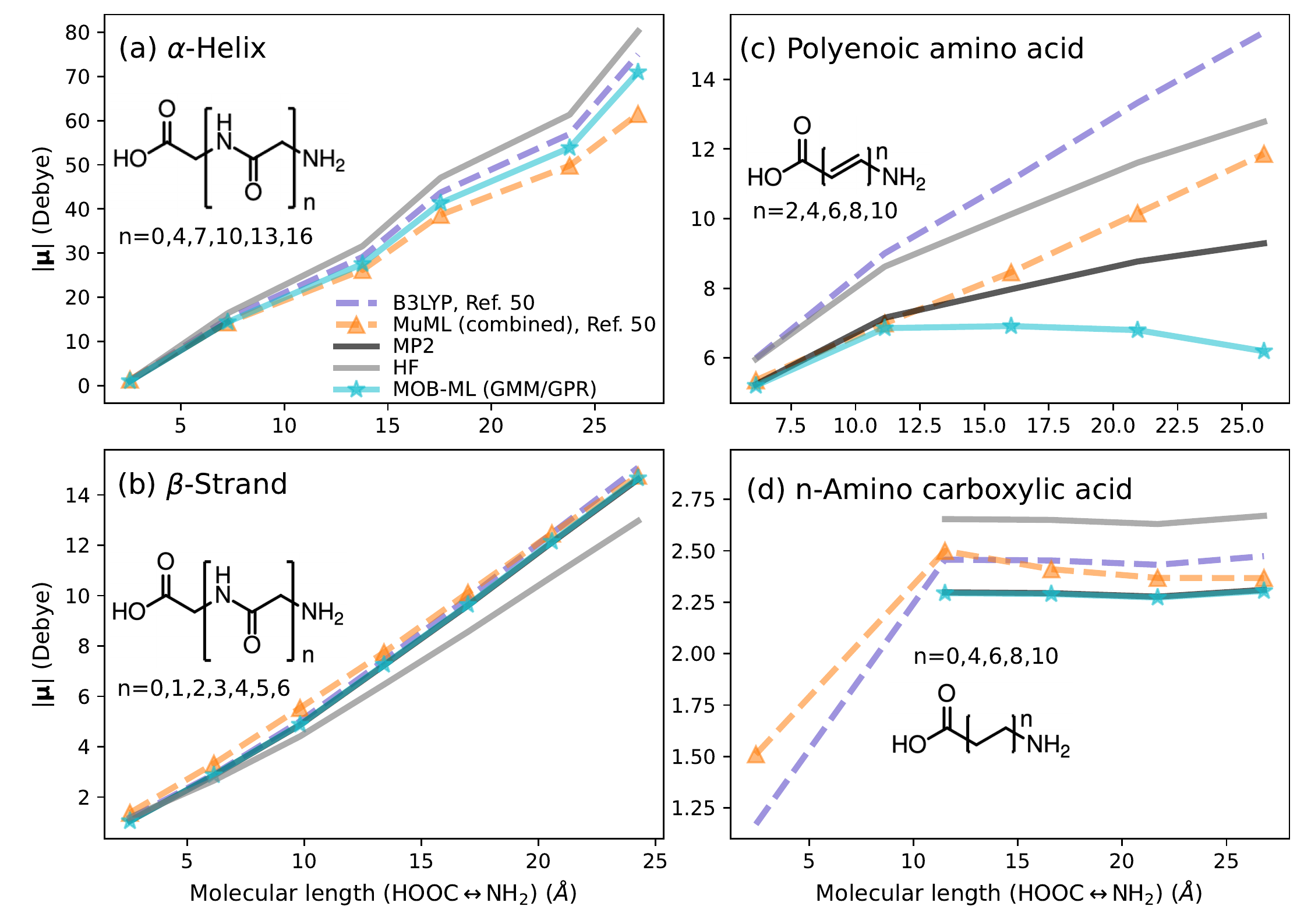}
    \caption{Dipole moment predictions for series of (a) $\alpha$-helix, (b) $\beta$-strand, (c) polyenoic amino acid, and (d) n-amino carboxylic acid using MOB-ML and MuML. The dipole moments of different molecules are plotted versus the chain length. The MOB-ML dipole moments are computed from the best MOB-ML(GMM/GPR) trained on 110,000 QM9 molecules. The results of MuML (combined) are predicted from the MuML model trained on 5400 QM7b molecules and extracted from Fig. 6 in Ref.~\citenum{veit2020predicting}. The reference dipole moments computed using MP2/cc-pVTZ (for MOB-ML) and B3LYP/daDZ (for MuML) are shown in the plots, and the HF dipole moments are also provided for further discussion. In (a), the molecules with $n=7,10,13,16$ cannot be computed by MP2/cc-pVTZ, and results from other theories and ML models are shown.}    
    \label{fig:challenge_dataset}
\end{figure*}
Figure~\ref{fig:timing} displays the accuracy improvements of test QM9 molecules as functions of training costs using MOB-ML (GPR) and MOB-ML (GMM/GPR) for dipole moments and energies. 
These models are collected on 8 NVIDIA Tesla V100-SXM2-32GB GPUs. 
Since each local GPR could be regressed independently on different GPUs, GMM/GPR is a highly-parallelized approach with excellent multi-GPU speedups.
For both dipole and energy learning, it is clear that GMM/GPR provides much lower training costs compared with learning without clustering.
Across all the training sizes that we could train directly with GPR, GMM/GPR could achieve more than 68.5 and 21.4 times speedups for dipole and energy learning, respectively, without loss of accuracy and transferability. 
For example, the best GMM/GPR energy model training on 110,000 molecules only takes 20.7 hrs, but it is over 7 times more expensive than the best dipole model (146.5 hrs). 
This is because the training points of dipole moment are 3 times larger than ones of energy for each molecule.

\subsection{Predictions of dipole moments of four challenge cases}
To illustrate the accuracy of MOB-ML in the actual biochemical systems, we further assess the prediction accuracy of the best MOB-ML (GMM/GPR) model on four different sets of peptides, termed as "challenging dataset". 
This challenging dataset is firstly introduced in Veit et al. \cite{veit2020predicting}, and in this study, we also included the literature results predicted by MuML model \cite{veit2020predicting}. 
All the true and predicted dipole moments from different theories and ML models are plotted as a function of the chain length in Fig.~\ref{fig:challenge_dataset}.
Table S3 in SI summarizes the predicted and true dipole moments and energies using the best QM9 energy model (GMM/GPR trained on 110,000 molecules). 
We note that MP2/cc-pVTZ is still an affordable theory for the QM9 benchmark dataset, but it is nearly impossible to obtain MP2 energy for the molecules with more than 30 heavy atoms, for instance, $\alpha$-helix molecules with $n=7,10,13,16$ in the challenging dataset. 
Therefore, no true MP2/cc-pVTZ results are provided for these large $\alpha$-helix molecules; meanwhile, MOB-ML provides reasonable dipole moment predictions for these molecules. 

Except the large polyenonic amino acids in Fig.~\ref{fig:challenge_dataset}(c), MOB-ML provides nearly identical predicted dipole moments as MP2 for all other molecules in panels (a), (b), and (d), which indicates that the MOB-ML model for dipole moments has an excellent transferability to large molecules.
MuML (combined) model also has a deteriorated accuracy for all the molecules in polyenoic amino acids.
In this case, the MOB-ML model over-corrects the results from HF calculation, and the size of the errors increases dramatically with the increasing of sizes. 
The major contributors to the total dipole moments for both cases are the local polarization of the end groups, and therefore the total dipole moments should be nearly constant predictions with the increase of the molecular length. 
This observation agrees with the ML predictions and the true dipole moments of n-amino carboxylic acids, but not the true values of polyenoic amino acids.
Veit et al. \cite{veit2020predicting} assess the partial charges for each functional group carefully to understand the puzzling results of polyenoic amino acids. 
The partial charge distribution suggests that end groups in polyenoic amino acids carry a net positive and negative charge, leading to an increase of dipole moments with increasing molecular length. 

\section{Conclusion}
In this study, we extend the MOB-ML framework to learn pairwise contributions of electron correlation part of dipole moments accurately and transferablely using the information computed from HF calculations. 
The feature derivatives are appended to the original energy features to form the dipole feature set to include more information in the MOB representation for dipole prediction.
The introduction of GPR with derivatives algorithm leads to efficient and physical modeling of dipole moments by satisfying the significant properties of equivariance and ML response. For water and other small molecules, MOB-ML could provide more accurate predictions for the dipole moments and energies by learning two tasks separately than simultaneously. To generate a universal dipole model and energy model for organic molecules, we apply MOB-ML to the QM9 dataset and train the two sets of labels separately using the corresponding GPR training protocols. 
The GMM/GPR framework that constructs local scalable GPRs for clusters detected by GMM clustering approach is also applied to reduce the learning costs of MOB-ML and scale up its training to 110,000 molecules without loss of accuracy. MOB-ML (GMM/GPR) model could achieve accuracies of 4.21 mDebye and 0.045 kcal/mol by learning 110000 QM9 molecules for dipole moments and energies, respectively, and an accuracy of 0.99 kcal/mol by only training on 100 QM9 molecules for energies.
MOB-ML is superior to all other literature results for both dipole moments and molecular energies and provides accurate and transferable results to most of the tested peptides with different three-dimensional structures. 
As a future study direction, this MOB feature design and GPR learning framework could be extended to model other response properties, such as polarizability and excited state properties. Furthermore, by taking advantage of the existence of the currently best MP2/cc-pVTZ models, multi-fidelity learning or $\Delta$-learning \cite{ramakrishnan2015big} approaches could also be applied to learn the MOB-ML model at CCSD(T)/cc-pVTZ level by learning the differences between CCSD(T) and MP2 values.

\begin{acknowledgments}
We thank Vignesh Bhethanabotla for his help in improving the quality of this manuscript. TFM acknowledges support from the US Army Research Laboratory (W911NF-12-2-0023), the US Department of Energy (DE-SC0019390), the Caltech DeLogi Fund, and the Camille and Henry Dreyfus Foundation (Award ML-20-196). Computational resources were provided by the National Energy Research Scientific Computing Center (NERSC), a DOE Office of Science User Facility supported by the DOE Office of Science under contract DE-AC02-05CH11231.
\end{acknowledgments}

\section*{Supporting Information}
The reference pairwise decomposed dipole moments and energies and MOB features of QM7b-T, QM9, and GDB-13-T are available at Caltech Data: \url{https://data.caltech.edu/records/1177}. The corresponding HF dipole moments and energies are also available. The total dipole moment and energy data of challenging datasets are also included in this online dataset. The implementation of the multi-GPU AltBBMM and GMM are available online at \url{https://github.com/SUSYUSTC/BBMM.git}. The parameters of training the models are included in the Supporting information. Table S1, S2, and S3 list the MOB-ML results plotted in Fig. ~\ref{fig:water_lc}, ~\ref{fig:qm9_to_qm9}, and ~\ref{fig:challenge_dataset}, respectively.

\bibliography{main}

\begin{thebibliography}{83}%
\makeatletter
\providecommand \@ifxundefined [1]{%
 \@ifx{#1\undefined}
}%
\providecommand \@ifnum [1]{%
 \ifnum #1\expandafter \@firstoftwo
 \else \expandafter \@secondoftwo
 \fi
}%
\providecommand \@ifx [1]{%
 \ifx #1\expandafter \@firstoftwo
 \else \expandafter \@secondoftwo
 \fi
}%
\providecommand \natexlab [1]{#1}%
\providecommand \enquote  [1]{``#1''}%
\providecommand \bibnamefont  [1]{#1}%
\providecommand \bibfnamefont [1]{#1}%
\providecommand \citenamefont [1]{#1}%
\providecommand \href@noop [0]{\@secondoftwo}%
\providecommand \href [0]{\begingroup \@sanitize@url \@href}%
\providecommand \@href[1]{\@@startlink{#1}\@@href}%
\providecommand \@@href[1]{\endgroup#1\@@endlink}%
\providecommand \@sanitize@url [0]{\catcode `\\12\catcode `\$12\catcode
  `\&12\catcode `\#12\catcode `\^12\catcode `\_12\catcode `\%12\relax}%
\providecommand \@@startlink[1]{}%
\providecommand \@@endlink[0]{}%
\providecommand \url  [0]{\begingroup\@sanitize@url \@url }%
\providecommand \@url [1]{\endgroup\@href {#1}{\urlprefix }}%
\providecommand \urlprefix  [0]{URL }%
\providecommand \Eprint [0]{\href }%
\providecommand \doibase [0]{https://doi.org/}%
\providecommand \selectlanguage [0]{\@gobble}%
\providecommand \bibinfo  [0]{\@secondoftwo}%
\providecommand \bibfield  [0]{\@secondoftwo}%
\providecommand \translation [1]{[#1]}%
\providecommand \BibitemOpen [0]{}%
\providecommand \bibitemStop [0]{}%
\providecommand \bibitemNoStop [0]{.\EOS\space}%
\providecommand \EOS [0]{\spacefactor3000\relax}%
\providecommand \BibitemShut  [1]{\csname bibitem#1\endcsname}%
\let\auto@bib@innerbib\@empty
\bibitem [{\citenamefont {von Lilienfeld}\ and\ \citenamefont
  {Burke}(2020)}]{von2020retrospective}%
  \BibitemOpen
  \bibfield  {author} {\bibinfo {author} {\bibfnamefont {O.~A.}\ \bibnamefont
  {von Lilienfeld}}\ and\ \bibinfo {author} {\bibfnamefont {K.}~\bibnamefont
  {Burke}},\ }\bibfield  {title} {\enquote {\bibinfo {title} {Retrospective on
  a decade of machine learning for chemical discovery},}\ }\href@noop {}
  {\bibfield  {journal} {\bibinfo  {journal} {Nat. Commun.}\ }\textbf {\bibinfo
  {volume} {11}},\ \bibinfo {pages} {1--4} (\bibinfo {year}
  {2020})}\BibitemShut {NoStop}%
\bibitem [{\citenamefont {Westermayr}\ \emph {et~al.}(2021)\citenamefont
  {Westermayr}, \citenamefont {Gastegger}, \citenamefont {Sch{\"u}tt},\ and\
  \citenamefont {Maurer}}]{westermayr2021perspective}%
  \BibitemOpen
  \bibfield  {author} {\bibinfo {author} {\bibfnamefont {J.}~\bibnamefont
  {Westermayr}}, \bibinfo {author} {\bibfnamefont {M.}~\bibnamefont
  {Gastegger}}, \bibinfo {author} {\bibfnamefont {K.~T.}\ \bibnamefont
  {Sch{\"u}tt}},\ and\ \bibinfo {author} {\bibfnamefont {R.~J.}\ \bibnamefont
  {Maurer}},\ }\bibfield  {title} {\enquote {\bibinfo {title} {Perspective on
  integrating machine learning into computational chemistry and materials
  science},}\ }\href@noop {} {\bibfield  {journal} {\bibinfo  {journal} {J.
  Chem. Phys.}\ }\textbf {\bibinfo {volume} {154}},\ \bibinfo {pages} {230903}
  (\bibinfo {year} {2021})}\BibitemShut {NoStop}%
\bibitem [{\citenamefont {Freeze}, \citenamefont {Kelly},\ and\ \citenamefont
  {Batista}(2019)}]{freeze2019search}%
  \BibitemOpen
  \bibfield  {author} {\bibinfo {author} {\bibfnamefont {J.~G.}\ \bibnamefont
  {Freeze}}, \bibinfo {author} {\bibfnamefont {H.~R.}\ \bibnamefont {Kelly}},\
  and\ \bibinfo {author} {\bibfnamefont {V.~S.}\ \bibnamefont {Batista}},\
  }\bibfield  {title} {\enquote {\bibinfo {title} {Search for catalysts by
  inverse design: artificial intelligence, mountain climbers, and
  alchemists},}\ }\href@noop {} {\bibfield  {journal} {\bibinfo  {journal}
  {Chem. Rev.}\ }\textbf {\bibinfo {volume} {119}},\ \bibinfo {pages}
  {6595--6612} (\bibinfo {year} {2019})}\BibitemShut {NoStop}%
\bibitem [{\citenamefont {Elton}\ \emph {et~al.}(2019)\citenamefont {Elton},
  \citenamefont {Boukouvalas}, \citenamefont {Fuge},\ and\ \citenamefont
  {Chung}}]{elton2019deep}%
  \BibitemOpen
  \bibfield  {author} {\bibinfo {author} {\bibfnamefont {D.~C.}\ \bibnamefont
  {Elton}}, \bibinfo {author} {\bibfnamefont {Z.}~\bibnamefont {Boukouvalas}},
  \bibinfo {author} {\bibfnamefont {M.~D.}\ \bibnamefont {Fuge}},\ and\
  \bibinfo {author} {\bibfnamefont {P.~W.}\ \bibnamefont {Chung}},\ }\bibfield
  {title} {\enquote {\bibinfo {title} {Deep learning for molecular design—a
  review of the state of the art},}\ }\href@noop {} {\bibfield  {journal}
  {\bibinfo  {journal} {Mol. Syst. Des. Eng.}\ }\textbf {\bibinfo {volume}
  {4}},\ \bibinfo {pages} {828--849} (\bibinfo {year} {2019})}\BibitemShut
  {NoStop}%
\bibitem [{\citenamefont {von Lilienfeld}, \citenamefont {Lins},\ and\
  \citenamefont {Rothlisberger}(2005)}]{von2005variational}%
  \BibitemOpen
  \bibfield  {author} {\bibinfo {author} {\bibfnamefont {O.~A.}\ \bibnamefont
  {von Lilienfeld}}, \bibinfo {author} {\bibfnamefont {R.~D.}\ \bibnamefont
  {Lins}},\ and\ \bibinfo {author} {\bibfnamefont {U.}~\bibnamefont
  {Rothlisberger}},\ }\bibfield  {title} {\enquote {\bibinfo {title}
  {Variational particle number approach for rational compound design},}\
  }\href@noop {} {\bibfield  {journal} {\bibinfo  {journal} {Phys. Rev. Lett.}\
  }\textbf {\bibinfo {volume} {95}},\ \bibinfo {pages} {153002} (\bibinfo
  {year} {2005})}\BibitemShut {NoStop}%
\bibitem [{\citenamefont {Von~Lilienfeld}\ and\ \citenamefont
  {Tuckerman}(2007)}]{von2007alchemical}%
  \BibitemOpen
  \bibfield  {author} {\bibinfo {author} {\bibfnamefont {O.~A.}\ \bibnamefont
  {Von~Lilienfeld}}\ and\ \bibinfo {author} {\bibfnamefont {M.}~\bibnamefont
  {Tuckerman}},\ }\bibfield  {title} {\enquote {\bibinfo {title} {Alchemical
  variations of intermolecular energies according to molecular grand-canonical
  ensemble density functional theory},}\ }\href@noop {} {\bibfield  {journal}
  {\bibinfo  {journal} {J. Chem. Theory Comput.}\ }\textbf {\bibinfo {volume}
  {3}},\ \bibinfo {pages} {1083--1090} (\bibinfo {year} {2007})}\BibitemShut
  {NoStop}%
\bibitem [{\citenamefont {Popova}, \citenamefont {Isayev},\ and\ \citenamefont
  {Tropsha}(2018)}]{Popova2018}%
  \BibitemOpen
  \bibfield  {author} {\bibinfo {author} {\bibfnamefont {M.}~\bibnamefont
  {Popova}}, \bibinfo {author} {\bibfnamefont {O.}~\bibnamefont {Isayev}},\
  and\ \bibinfo {author} {\bibfnamefont {A.}~\bibnamefont {Tropsha}},\
  }\bibfield  {title} {\enquote {\bibinfo {title} {{Deep reinforcement learning
  for de novo drug design}},}\ }\href {https://doi.org/10.1126/sciadv.aap7885}
  {\bibfield  {journal} {\bibinfo  {journal} {Sci. Adv.}\ }\textbf {\bibinfo
  {volume} {4}},\ \bibinfo {pages} {eaap7885} (\bibinfo {year}
  {2018})}\BibitemShut {NoStop}%
\bibitem [{\citenamefont {Yang}\ \emph {et~al.}(2019)\citenamefont {Yang},
  \citenamefont {Wang}, \citenamefont {Byrne}, \citenamefont {Schneider},\ and\
  \citenamefont {Yang}}]{yang2019concepts}%
  \BibitemOpen
  \bibfield  {author} {\bibinfo {author} {\bibfnamefont {X.}~\bibnamefont
  {Yang}}, \bibinfo {author} {\bibfnamefont {Y.}~\bibnamefont {Wang}}, \bibinfo
  {author} {\bibfnamefont {R.}~\bibnamefont {Byrne}}, \bibinfo {author}
  {\bibfnamefont {G.}~\bibnamefont {Schneider}},\ and\ \bibinfo {author}
  {\bibfnamefont {S.}~\bibnamefont {Yang}},\ }\bibfield  {title} {\enquote
  {\bibinfo {title} {Concepts of artificial intelligence for computer-assisted
  drug discovery},}\ }\href@noop {} {\bibfield  {journal} {\bibinfo  {journal}
  {Chem. Rev.}\ }\textbf {\bibinfo {volume} {119}},\ \bibinfo {pages}
  {10520--10594} (\bibinfo {year} {2019})}\BibitemShut {NoStop}%
\bibitem [{\citenamefont {Oganov}\ \emph {et~al.}(2019)\citenamefont {Oganov},
  \citenamefont {Pickard}, \citenamefont {Zhu},\ and\ \citenamefont
  {Needs}}]{oganov2019structure}%
  \BibitemOpen
  \bibfield  {author} {\bibinfo {author} {\bibfnamefont {A.~R.}\ \bibnamefont
  {Oganov}}, \bibinfo {author} {\bibfnamefont {C.~J.}\ \bibnamefont {Pickard}},
  \bibinfo {author} {\bibfnamefont {Q.}~\bibnamefont {Zhu}},\ and\ \bibinfo
  {author} {\bibfnamefont {R.~J.}\ \bibnamefont {Needs}},\ }\bibfield  {title}
  {\enquote {\bibinfo {title} {Structure prediction drives materials
  discovery},}\ }\href@noop {} {\bibfield  {journal} {\bibinfo  {journal} {Nat.
  Rev. Mater.}\ }\textbf {\bibinfo {volume} {4}},\ \bibinfo {pages} {331--348}
  (\bibinfo {year} {2019})}\BibitemShut {NoStop}%
\bibitem [{\citenamefont {Tran}\ \emph {et~al.}(2020)\citenamefont {Tran},
  \citenamefont {Tranchida}, \citenamefont {Wildey},\ and\ \citenamefont
  {Thompson}}]{tran2020multi}%
  \BibitemOpen
  \bibfield  {author} {\bibinfo {author} {\bibfnamefont {A.}~\bibnamefont
  {Tran}}, \bibinfo {author} {\bibfnamefont {J.}~\bibnamefont {Tranchida}},
  \bibinfo {author} {\bibfnamefont {T.}~\bibnamefont {Wildey}},\ and\ \bibinfo
  {author} {\bibfnamefont {A.~P.}\ \bibnamefont {Thompson}},\ }\bibfield
  {title} {\enquote {\bibinfo {title} {Multi-fidelity machine-learning with
  uncertainty quantification and bayesian optimization for materials design:
  Application to ternary random alloys},}\ }\href@noop {} {\bibfield  {journal}
  {\bibinfo  {journal} {J. Chem. Phys.}\ }\textbf {\bibinfo {volume} {153}},\
  \bibinfo {pages} {074705} (\bibinfo {year} {2020})}\BibitemShut {NoStop}%
\bibitem [{\citenamefont {Grisafi}\ \emph {et~al.}(2019)\citenamefont
  {Grisafi}, \citenamefont {Fabrizio}, \citenamefont {Meyer}, \citenamefont
  {Wilkins}, \citenamefont {Corminboeuf},\ and\ \citenamefont
  {Ceriotti}}]{Grisafi2018}%
  \BibitemOpen
  \bibfield  {author} {\bibinfo {author} {\bibfnamefont {A.}~\bibnamefont
  {Grisafi}}, \bibinfo {author} {\bibfnamefont {A.}~\bibnamefont {Fabrizio}},
  \bibinfo {author} {\bibfnamefont {B.}~\bibnamefont {Meyer}}, \bibinfo
  {author} {\bibfnamefont {D.~M.}\ \bibnamefont {Wilkins}}, \bibinfo {author}
  {\bibfnamefont {C.}~\bibnamefont {Corminboeuf}},\ and\ \bibinfo {author}
  {\bibfnamefont {M.}~\bibnamefont {Ceriotti}},\ }\bibfield  {title} {\enquote
  {\bibinfo {title} {Transferable machine-learning model of the electron
  density},}\ }\href@noop {} {\bibfield  {journal} {\bibinfo  {journal} {ACS
  Cent. Sci.}\ }\textbf {\bibinfo {volume} {5}},\ \bibinfo {pages} {57--64}
  (\bibinfo {year} {2019})}\BibitemShut {NoStop}%
\bibitem [{\citenamefont {Manzhos}\ and\ \citenamefont
  {Carrington~Jr}(2020)}]{manzhos2020neural}%
  \BibitemOpen
  \bibfield  {author} {\bibinfo {author} {\bibfnamefont {S.}~\bibnamefont
  {Manzhos}}\ and\ \bibinfo {author} {\bibfnamefont {T.}~\bibnamefont
  {Carrington~Jr}},\ }\bibfield  {title} {\enquote {\bibinfo {title} {Neural
  network potential energy surfaces for small molecules and reactions},}\
  }\href@noop {} {\bibfield  {journal} {\bibinfo  {journal} {Chem. Rev.}\
  }\textbf {\bibinfo {volume} {121}},\ \bibinfo {pages} {10187--10217}
  (\bibinfo {year} {2020})}\BibitemShut {NoStop}%
\bibitem [{\citenamefont {Ceriotti}(2019)}]{ceriotti2019unsupervised}%
  \BibitemOpen
  \bibfield  {author} {\bibinfo {author} {\bibfnamefont {M.}~\bibnamefont
  {Ceriotti}},\ }\bibfield  {title} {\enquote {\bibinfo {title} {Unsupervised
  machine learning in atomistic simulations, between predictions and
  understanding},}\ }\href@noop {} {\bibfield  {journal} {\bibinfo  {journal}
  {J. Chem. Phys.}\ }\textbf {\bibinfo {volume} {150}},\ \bibinfo {pages}
  {150901} (\bibinfo {year} {2019})}\BibitemShut {NoStop}%
\bibitem [{\citenamefont {Tshitoyan}\ \emph {et~al.}(2019)\citenamefont
  {Tshitoyan}, \citenamefont {Dagdelen}, \citenamefont {Weston}, \citenamefont
  {Dunn}, \citenamefont {Rong}, \citenamefont {Kononova}, \citenamefont
  {Persson}, \citenamefont {Ceder},\ and\ \citenamefont
  {Jain}}]{tshitoyan2019unsupervised}%
  \BibitemOpen
  \bibfield  {author} {\bibinfo {author} {\bibfnamefont {V.}~\bibnamefont
  {Tshitoyan}}, \bibinfo {author} {\bibfnamefont {J.}~\bibnamefont {Dagdelen}},
  \bibinfo {author} {\bibfnamefont {L.}~\bibnamefont {Weston}}, \bibinfo
  {author} {\bibfnamefont {A.}~\bibnamefont {Dunn}}, \bibinfo {author}
  {\bibfnamefont {Z.}~\bibnamefont {Rong}}, \bibinfo {author} {\bibfnamefont
  {O.}~\bibnamefont {Kononova}}, \bibinfo {author} {\bibfnamefont {K.~A.}\
  \bibnamefont {Persson}}, \bibinfo {author} {\bibfnamefont {G.}~\bibnamefont
  {Ceder}},\ and\ \bibinfo {author} {\bibfnamefont {A.}~\bibnamefont {Jain}},\
  }\bibfield  {title} {\enquote {\bibinfo {title} {Unsupervised word embeddings
  capture latent knowledge from materials science literature},}\ }\href@noop {}
  {\bibfield  {journal} {\bibinfo  {journal} {Nature}\ }\textbf {\bibinfo
  {volume} {571}},\ \bibinfo {pages} {95--98} (\bibinfo {year}
  {2019})}\BibitemShut {NoStop}%
\bibitem [{\citenamefont {Zhou}\ \emph {et~al.}(2019)\citenamefont {Zhou},
  \citenamefont {Kearnes}, \citenamefont {Li}, \citenamefont {Zare},\ and\
  \citenamefont {Riley}}]{zhou2019optimization}%
  \BibitemOpen
  \bibfield  {author} {\bibinfo {author} {\bibfnamefont {Z.}~\bibnamefont
  {Zhou}}, \bibinfo {author} {\bibfnamefont {S.}~\bibnamefont {Kearnes}},
  \bibinfo {author} {\bibfnamefont {L.}~\bibnamefont {Li}}, \bibinfo {author}
  {\bibfnamefont {R.~N.}\ \bibnamefont {Zare}},\ and\ \bibinfo {author}
  {\bibfnamefont {P.}~\bibnamefont {Riley}},\ }\bibfield  {title} {\enquote
  {\bibinfo {title} {Optimization of molecules via deep reinforcement
  learning},}\ }\href@noop {} {\bibfield  {journal} {\bibinfo  {journal} {Sci.
  Rep.}\ }\textbf {\bibinfo {volume} {9}},\ \bibinfo {pages} {1--10} (\bibinfo
  {year} {2019})}\BibitemShut {NoStop}%
\bibitem [{\citenamefont {Sanchez-Lengeling}\ and\ \citenamefont
  {Aspuru-Guzik}(2018)}]{Sanchez-Lengeling2018}%
  \BibitemOpen
  \bibfield  {author} {\bibinfo {author} {\bibfnamefont {B.}~\bibnamefont
  {Sanchez-Lengeling}}\ and\ \bibinfo {author} {\bibfnamefont {A.}~\bibnamefont
  {Aspuru-Guzik}},\ }\bibfield  {title} {\enquote {\bibinfo {title} {{Inverse
  molecular design using machine learning: Generative models for matter
  engineering.}}}\ }\href {https://doi.org/10.1126/science.aat2663} {\bibfield
  {journal} {\bibinfo  {journal} {Science}\ }\textbf {\bibinfo {volume}
  {361}},\ \bibinfo {pages} {360--365} (\bibinfo {year} {2018})}\BibitemShut
  {NoStop}%
\bibitem [{\citenamefont {Schwalbe-Koda}\ and\ \citenamefont
  {G{\'o}mez-Bombarelli}(2020)}]{schwalbe2020generative}%
  \BibitemOpen
  \bibfield  {author} {\bibinfo {author} {\bibfnamefont {D.}~\bibnamefont
  {Schwalbe-Koda}}\ and\ \bibinfo {author} {\bibfnamefont {R.}~\bibnamefont
  {G{\'o}mez-Bombarelli}},\ }\bibfield  {title} {\enquote {\bibinfo {title}
  {Generative models for automatic chemical design},}\ }in\ \href@noop {}
  {\emph {\bibinfo {booktitle} {Machine Learning Meets Quantum Physics}}}\
  (\bibinfo  {publisher} {Springer},\ \bibinfo {year} {2020})\ pp.\ \bibinfo
  {pages} {445--467}\BibitemShut {NoStop}%
\bibitem [{\citenamefont {Bart{\'{o}}k}\ \emph {et~al.}(2010)\citenamefont
  {Bart{\'{o}}k}, \citenamefont {Payne}, \citenamefont {Kondor},\ and\
  \citenamefont {Cs{\'{a}}nyi}}]{Bartok2010}%
  \BibitemOpen
  \bibfield  {author} {\bibinfo {author} {\bibfnamefont {A.~P.}\ \bibnamefont
  {Bart{\'{o}}k}}, \bibinfo {author} {\bibfnamefont {M.~C.}\ \bibnamefont
  {Payne}}, \bibinfo {author} {\bibfnamefont {R.}~\bibnamefont {Kondor}},\ and\
  \bibinfo {author} {\bibfnamefont {G.}~\bibnamefont {Cs{\'{a}}nyi}},\
  }\bibfield  {title} {\enquote {\bibinfo {title} {{Gaussian approximation
  potentials: The accuracy of quantum mechanics, without the electrons}},}\
  }\href {https://doi.org/10.1103/PhysRevLett.104.136403} {\bibfield  {journal}
  {\bibinfo  {journal} {Phys. Rev. Lett.}\ }\textbf {\bibinfo {volume} {104}},\
  \bibinfo {pages} {136403} (\bibinfo {year} {2010})}\BibitemShut {NoStop}%
\bibitem [{\citenamefont {Rupp}\ \emph {et~al.}(2012)\citenamefont {Rupp},
  \citenamefont {Tkatchenko}, \citenamefont {M{\"{u}}ller},\ and\ \citenamefont
  {{von Lilienfeld}}}]{rupp2012fast}%
  \BibitemOpen
  \bibfield  {author} {\bibinfo {author} {\bibfnamefont {M.}~\bibnamefont
  {Rupp}}, \bibinfo {author} {\bibfnamefont {A.}~\bibnamefont {Tkatchenko}},
  \bibinfo {author} {\bibfnamefont {K.-R.}\ \bibnamefont {M{\"{u}}ller}},\ and\
  \bibinfo {author} {\bibfnamefont {O.~A.}\ \bibnamefont {{von Lilienfeld}}},\
  }\bibfield  {title} {\enquote {\bibinfo {title} {{Fast and accurate modeling
  of molecular atomization energies with machine learning}},}\ }\href@noop {}
  {\bibfield  {journal} {\bibinfo  {journal} {Phys. Rev. Lett.}\ }\textbf
  {\bibinfo {volume} {108}},\ \bibinfo {pages} {58301} (\bibinfo {year}
  {2012})}\BibitemShut {NoStop}%
\bibitem [{\citenamefont {Montavon}\ \emph {et~al.}(2013)\citenamefont
  {Montavon}, \citenamefont {Rupp}, \citenamefont {Gobre}, \citenamefont
  {Vazquez-Mayagoitia}, \citenamefont {Hansen}, \citenamefont {Tkatchenko},
  \citenamefont {M{\"{u}}ller},\ and\ \citenamefont {{von
  Lilienfeld}}}]{VonLilienfeld2013}%
  \BibitemOpen
  \bibfield  {author} {\bibinfo {author} {\bibfnamefont {G.}~\bibnamefont
  {Montavon}}, \bibinfo {author} {\bibfnamefont {M.}~\bibnamefont {Rupp}},
  \bibinfo {author} {\bibfnamefont {V.}~\bibnamefont {Gobre}}, \bibinfo
  {author} {\bibfnamefont {A.}~\bibnamefont {Vazquez-Mayagoitia}}, \bibinfo
  {author} {\bibfnamefont {K.}~\bibnamefont {Hansen}}, \bibinfo {author}
  {\bibfnamefont {A.}~\bibnamefont {Tkatchenko}}, \bibinfo {author}
  {\bibfnamefont {K.-R.}\ \bibnamefont {M{\"{u}}ller}},\ and\ \bibinfo {author}
  {\bibfnamefont {O.~A.}\ \bibnamefont {{von Lilienfeld}}},\ }\bibfield
  {title} {\enquote {\bibinfo {title} {{Machine learning of molecular
  electronic properties in chemical compound space}},}\ }\href@noop {}
  {\bibfield  {journal} {\bibinfo  {journal} {New J. Phys.}\ }\textbf {\bibinfo
  {volume} {15}},\ \bibinfo {pages} {95003} (\bibinfo {year}
  {2013})}\BibitemShut {NoStop}%
\bibitem [{\citenamefont {Hansen}\ \emph {et~al.}(2013)\citenamefont {Hansen},
  \citenamefont {Montavon}, \citenamefont {Biegler}, \citenamefont {Fazli},
  \citenamefont {Rupp}, \citenamefont {Scheffler}, \citenamefont {{von
  Lilienfeld}}, \citenamefont {Tkatchenko},\ and\ \citenamefont
  {M{\"{u}}ller}}]{hansen2013assessment}%
  \BibitemOpen
  \bibfield  {author} {\bibinfo {author} {\bibfnamefont {K.}~\bibnamefont
  {Hansen}}, \bibinfo {author} {\bibfnamefont {G.}~\bibnamefont {Montavon}},
  \bibinfo {author} {\bibfnamefont {F.}~\bibnamefont {Biegler}}, \bibinfo
  {author} {\bibfnamefont {S.}~\bibnamefont {Fazli}}, \bibinfo {author}
  {\bibfnamefont {M.}~\bibnamefont {Rupp}}, \bibinfo {author} {\bibfnamefont
  {M.}~\bibnamefont {Scheffler}}, \bibinfo {author} {\bibfnamefont {O.~A.}\
  \bibnamefont {{von Lilienfeld}}}, \bibinfo {author} {\bibfnamefont
  {A.}~\bibnamefont {Tkatchenko}},\ and\ \bibinfo {author} {\bibfnamefont
  {K.-R.}\ \bibnamefont {M{\"{u}}ller}},\ }\bibfield  {title} {\enquote
  {\bibinfo {title} {{Assessment and validation of machine learning methods for
  predicting molecular atomization energies}},}\ }\href@noop {} {\bibfield
  {journal} {\bibinfo  {journal} {J. Chem. Theory Comput.}\ }\textbf {\bibinfo
  {volume} {9}},\ \bibinfo {pages} {3404} (\bibinfo {year} {2013})}\BibitemShut
  {NoStop}%
\bibitem [{\citenamefont {Gasparotto}\ and\ \citenamefont
  {Ceriotti}(2014)}]{Ceriotti2014}%
  \BibitemOpen
  \bibfield  {author} {\bibinfo {author} {\bibfnamefont {P.}~\bibnamefont
  {Gasparotto}}\ and\ \bibinfo {author} {\bibfnamefont {M.}~\bibnamefont
  {Ceriotti}},\ }\bibfield  {title} {\enquote {\bibinfo {title} {{Recognizing
  molecular patterns by machine learning: An agnostic structural definition of
  the hydrogen bond}},}\ }\href@noop {} {\bibfield  {journal} {\bibinfo
  {journal} {J. Chem. Phys.}\ }\textbf {\bibinfo {volume} {141}},\ \bibinfo
  {pages} {174110} (\bibinfo {year} {2014})}\BibitemShut {NoStop}%
\bibitem [{\citenamefont {Ramakrishnan}\ \emph {et~al.}(2015)\citenamefont
  {Ramakrishnan}, \citenamefont {Dral}, \citenamefont {Rupp},\ and\
  \citenamefont {{von Lilienfeld}}}]{ramakrishnan2015big}%
  \BibitemOpen
  \bibfield  {author} {\bibinfo {author} {\bibfnamefont {R.}~\bibnamefont
  {Ramakrishnan}}, \bibinfo {author} {\bibfnamefont {P.~O.}\ \bibnamefont
  {Dral}}, \bibinfo {author} {\bibfnamefont {M.}~\bibnamefont {Rupp}},\ and\
  \bibinfo {author} {\bibfnamefont {O.~A.}\ \bibnamefont {{von Lilienfeld}}},\
  }\bibfield  {title} {\enquote {\bibinfo {title} {{Big data meets quantum
  chemistry approximations: The $\Delta$-machine learning approach}},}\
  }\href@noop {} {\bibfield  {journal} {\bibinfo  {journal} {J. Chem. Theory
  Comput.}\ }\textbf {\bibinfo {volume} {11}},\ \bibinfo {pages} {2087}
  (\bibinfo {year} {2015})}\BibitemShut {NoStop}%
\bibitem [{\citenamefont {Brockherde}\ \emph {et~al.}(2017)\citenamefont
  {Brockherde}, \citenamefont {Vogt}, \citenamefont {Li}, \citenamefont
  {Tuckerman}, \citenamefont {Burke},\ and\ \citenamefont
  {M{\"{u}}ller}}]{Tuckerman}%
  \BibitemOpen
  \bibfield  {author} {\bibinfo {author} {\bibfnamefont {F.}~\bibnamefont
  {Brockherde}}, \bibinfo {author} {\bibfnamefont {L.}~\bibnamefont {Vogt}},
  \bibinfo {author} {\bibfnamefont {L.}~\bibnamefont {Li}}, \bibinfo {author}
  {\bibfnamefont {M.~E.}\ \bibnamefont {Tuckerman}}, \bibinfo {author}
  {\bibfnamefont {K.}~\bibnamefont {Burke}},\ and\ \bibinfo {author}
  {\bibfnamefont {K.-R.}\ \bibnamefont {M{\"{u}}ller}},\ }\bibfield  {title}
  {\enquote {\bibinfo {title} {{Bypassing the Kohn-Sham equations with machine
  learning}},}\ }\href@noop {} {\bibfield  {journal} {\bibinfo  {journal} {Nat.
  Commun.}\ }\textbf {\bibinfo {volume} {8}},\ \bibinfo {pages} {872} (\bibinfo
  {year} {2017})}\BibitemShut {NoStop}%
\bibitem [{\citenamefont {Kearnes}\ \emph {et~al.}(2016)\citenamefont
  {Kearnes}, \citenamefont {McCloskey}, \citenamefont {Berndl}, \citenamefont
  {Pande},\ and\ \citenamefont {Riley}}]{kearnes2016molecular}%
  \BibitemOpen
  \bibfield  {author} {\bibinfo {author} {\bibfnamefont {S.}~\bibnamefont
  {Kearnes}}, \bibinfo {author} {\bibfnamefont {K.}~\bibnamefont {McCloskey}},
  \bibinfo {author} {\bibfnamefont {M.}~\bibnamefont {Berndl}}, \bibinfo
  {author} {\bibfnamefont {V.}~\bibnamefont {Pande}},\ and\ \bibinfo {author}
  {\bibfnamefont {P.}~\bibnamefont {Riley}},\ }\bibfield  {title} {\enquote
  {\bibinfo {title} {{Molecular graph convolutions: Moving beyond
  fingerprints}},}\ }\href@noop {} {\bibfield  {journal} {\bibinfo  {journal}
  {J. Comput. Aided Mol. Des.}\ }\textbf {\bibinfo {volume} {30}},\ \bibinfo
  {pages} {595} (\bibinfo {year} {2016})}\BibitemShut {NoStop}%
\bibitem [{\citenamefont {Paesani}(2016)}]{Paesani2016}%
  \BibitemOpen
  \bibfield  {author} {\bibinfo {author} {\bibfnamefont {F.}~\bibnamefont
  {Paesani}},\ }\bibfield  {title} {\enquote {\bibinfo {title} {{Getting the
  right answers for the right reasons: toward predictive molecular simulations
  of water with many-body potential energy functions}},}\ }\href
  {https://doi.org/10.1021/acs.accounts.6b00285} {\bibfield  {journal}
  {\bibinfo  {journal} {Acc. Chem. Res.}\ }\textbf {\bibinfo {volume} {49}},\
  \bibinfo {pages} {1844} (\bibinfo {year} {2016})}\BibitemShut {NoStop}%
\bibitem [{\citenamefont {Behler}(2016)}]{Behler2016}%
  \BibitemOpen
  \bibfield  {author} {\bibinfo {author} {\bibfnamefont {J.}~\bibnamefont
  {Behler}},\ }\bibfield  {title} {\enquote {\bibinfo {title} {{Perspective:
  Machine learning potentials for atomistic simulations}},}\ }\href
  {https://doi.org/10.1063/1.4966192} {\bibfield  {journal} {\bibinfo
  {journal} {J. Chem. Phys.}\ }\textbf {\bibinfo {volume} {145}},\ \bibinfo
  {pages} {170901} (\bibinfo {year} {2016})}\BibitemShut {NoStop}%
\bibitem [{\citenamefont {Sch{\"{u}}tt}\ \emph {et~al.}(2017)\citenamefont
  {Sch{\"{u}}tt}, \citenamefont {Arbabzadah}, \citenamefont {Chmiela},
  \citenamefont {M{\"{u}}ller},\ and\ \citenamefont
  {Tkatchenko}}]{schutt2017quantum}%
  \BibitemOpen
  \bibfield  {author} {\bibinfo {author} {\bibfnamefont {K.~T.}\ \bibnamefont
  {Sch{\"{u}}tt}}, \bibinfo {author} {\bibfnamefont {F.}~\bibnamefont
  {Arbabzadah}}, \bibinfo {author} {\bibfnamefont {S.}~\bibnamefont {Chmiela}},
  \bibinfo {author} {\bibfnamefont {K.-R.}\ \bibnamefont {M{\"{u}}ller}},\ and\
  \bibinfo {author} {\bibfnamefont {A.}~\bibnamefont {Tkatchenko}},\ }\bibfield
   {title} {\enquote {\bibinfo {title} {{Quantum-chemical insights from deep
  tensor neural networks}},}\ }\href@noop {} {\bibfield  {journal} {\bibinfo
  {journal} {Nat. Commun.}\ }\textbf {\bibinfo {volume} {8}},\ \bibinfo {pages}
  {13890} (\bibinfo {year} {2017})}\BibitemShut {NoStop}%
\bibitem [{\citenamefont {Sch{\"u}tt}\ \emph {et~al.}(2017)\citenamefont
  {Sch{\"u}tt}, \citenamefont {Kindermans}, \citenamefont {Sauceda~Felix},
  \citenamefont {Chmiela}, \citenamefont {Tkatchenko},\ and\ \citenamefont
  {M{\"u}ller}}]{schutt2017schnet}%
  \BibitemOpen
  \bibfield  {author} {\bibinfo {author} {\bibfnamefont {K.}~\bibnamefont
  {Sch{\"u}tt}}, \bibinfo {author} {\bibfnamefont {P.-J.}\ \bibnamefont
  {Kindermans}}, \bibinfo {author} {\bibfnamefont {H.~E.}\ \bibnamefont
  {Sauceda~Felix}}, \bibinfo {author} {\bibfnamefont {S.}~\bibnamefont
  {Chmiela}}, \bibinfo {author} {\bibfnamefont {A.}~\bibnamefont
  {Tkatchenko}},\ and\ \bibinfo {author} {\bibfnamefont {K.-R.}\ \bibnamefont
  {M{\"u}ller}},\ }\bibfield  {title} {\enquote {\bibinfo {title} {Schnet: A
  continuous-filter convolutional neural network for modeling quantum
  interactions},}\ }\href@noop {} {\bibfield  {journal} {\bibinfo  {journal}
  {Advances in neural information processing systems}\ }\textbf {\bibinfo
  {volume} {30}} (\bibinfo {year} {2017})}\BibitemShut {NoStop}%
\bibitem [{\citenamefont {Smith}, \citenamefont {Isayev},\ and\ \citenamefont
  {Roitberg}(2017)}]{Smith2017}%
  \BibitemOpen
  \bibfield  {author} {\bibinfo {author} {\bibfnamefont {J.~S.}\ \bibnamefont
  {Smith}}, \bibinfo {author} {\bibfnamefont {O.}~\bibnamefont {Isayev}},\ and\
  \bibinfo {author} {\bibfnamefont {A.~E.}\ \bibnamefont {Roitberg}},\
  }\bibfield  {title} {\enquote {\bibinfo {title} {{ANI-1: An extensible neural
  network potential with DFT accuracy at force field computational cost}},}\
  }\href {https://doi.org/10.1039/C6SC05720A} {\bibfield  {journal} {\bibinfo
  {journal} {Chem. Sci.}\ }\textbf {\bibinfo {volume} {8}},\ \bibinfo {pages}
  {3192--3203} (\bibinfo {year} {2017})}\BibitemShut {NoStop}%
\bibitem [{\citenamefont {Welborn}, \citenamefont {Cheng},\ and\ \citenamefont
  {{Miller III}}(2018)}]{Welborn2018}%
  \BibitemOpen
  \bibfield  {author} {\bibinfo {author} {\bibfnamefont {M.}~\bibnamefont
  {Welborn}}, \bibinfo {author} {\bibfnamefont {L.}~\bibnamefont {Cheng}},\
  and\ \bibinfo {author} {\bibfnamefont {T.~F.}\ \bibnamefont {{Miller III}}},\
  }\bibfield  {title} {\enquote {\bibinfo {title} {{Transferability in machine
  learning for electronic structure via the molecular orbital basis}},}\ }\href
  {https://doi.org/10.1021/acs.jctc.8b00636} {\bibfield  {journal} {\bibinfo
  {journal} {J. Chem. Theory Comput.}\ }\textbf {\bibinfo {volume} {14}},\
  \bibinfo {pages} {4772--4779} (\bibinfo {year} {2018})}\BibitemShut {NoStop}%
\bibitem [{\citenamefont {Wu}\ \emph {et~al.}(2018)\citenamefont {Wu},
  \citenamefont {Ramsundar}, \citenamefont {Feinberg}, \citenamefont {Gomes},
  \citenamefont {Geniesse}, \citenamefont {Pappu}, \citenamefont {Leswing},\
  and\ \citenamefont {Pande}}]{wu2018moleculenet}%
  \BibitemOpen
  \bibfield  {author} {\bibinfo {author} {\bibfnamefont {Z.}~\bibnamefont
  {Wu}}, \bibinfo {author} {\bibfnamefont {B.}~\bibnamefont {Ramsundar}},
  \bibinfo {author} {\bibfnamefont {E.~N.}\ \bibnamefont {Feinberg}}, \bibinfo
  {author} {\bibfnamefont {J.}~\bibnamefont {Gomes}}, \bibinfo {author}
  {\bibfnamefont {C.}~\bibnamefont {Geniesse}}, \bibinfo {author}
  {\bibfnamefont {A.~S.}\ \bibnamefont {Pappu}}, \bibinfo {author}
  {\bibfnamefont {K.}~\bibnamefont {Leswing}},\ and\ \bibinfo {author}
  {\bibfnamefont {V.}~\bibnamefont {Pande}},\ }\bibfield  {title} {\enquote
  {\bibinfo {title} {{MoleculeNet: A benchmark for molecular machine
  learning}},}\ }\href@noop {} {\bibfield  {journal} {\bibinfo  {journal}
  {Chem. Sci.}\ }\textbf {\bibinfo {volume} {9}},\ \bibinfo {pages} {513}
  (\bibinfo {year} {2018})}\BibitemShut {NoStop}%
\bibitem [{\citenamefont {Nguyen}\ \emph {et~al.}(2018)\citenamefont {Nguyen},
  \citenamefont {Sz{\'{e}}kely}, \citenamefont {Imbalzano}, \citenamefont
  {Behler}, \citenamefont {Cs{\'{a}}nyi}, \citenamefont {Ceriotti},
  \citenamefont {G{\"{o}}tz},\ and\ \citenamefont {Paesani}}]{Nguyen2018}%
  \BibitemOpen
  \bibfield  {author} {\bibinfo {author} {\bibfnamefont {T.~T.}\ \bibnamefont
  {Nguyen}}, \bibinfo {author} {\bibfnamefont {E.}~\bibnamefont
  {Sz{\'{e}}kely}}, \bibinfo {author} {\bibfnamefont {G.}~\bibnamefont
  {Imbalzano}}, \bibinfo {author} {\bibfnamefont {J.}~\bibnamefont {Behler}},
  \bibinfo {author} {\bibfnamefont {G.}~\bibnamefont {Cs{\'{a}}nyi}}, \bibinfo
  {author} {\bibfnamefont {M.}~\bibnamefont {Ceriotti}}, \bibinfo {author}
  {\bibfnamefont {A.~W.}\ \bibnamefont {G{\"{o}}tz}},\ and\ \bibinfo {author}
  {\bibfnamefont {F.}~\bibnamefont {Paesani}},\ }\bibfield  {title} {\enquote
  {\bibinfo {title} {{Comparison of permutationally invariant polynomials,
  neural networks, and Gaussian approximation potentials in representing water
  interactions through many-body expansions}},}\ }\href@noop {} {\bibfield
  {journal} {\bibinfo  {journal} {J. Chem. Phys.}\ }\textbf {\bibinfo {volume}
  {148}},\ \bibinfo {pages} {241725} (\bibinfo {year} {2018})}\BibitemShut
  {NoStop}%
\bibitem [{\citenamefont {Yao}\ \emph {et~al.}(2018)\citenamefont {Yao},
  \citenamefont {Herr}, \citenamefont {Toth}, \citenamefont {McKintyre},\ and\
  \citenamefont {Parkhill}}]{Yao2018}%
  \BibitemOpen
  \bibfield  {author} {\bibinfo {author} {\bibfnamefont {K.}~\bibnamefont
  {Yao}}, \bibinfo {author} {\bibfnamefont {J.~E.}\ \bibnamefont {Herr}},
  \bibinfo {author} {\bibfnamefont {D.~W.}\ \bibnamefont {Toth}}, \bibinfo
  {author} {\bibfnamefont {R.}~\bibnamefont {McKintyre}},\ and\ \bibinfo
  {author} {\bibfnamefont {J.}~\bibnamefont {Parkhill}},\ }\bibfield  {title}
  {\enquote {\bibinfo {title} {The tensormol-0.1 model chemistry: A neural
  network augmented with long-range physics},}\ }\href
  {https://doi.org/10.1039/c7sc04934j} {\bibfield  {journal} {\bibinfo
  {journal} {Chem. Sci.}\ }\textbf {\bibinfo {volume} {9}},\ \bibinfo {pages}
  {2261--2269} (\bibinfo {year} {2018})}\BibitemShut {NoStop}%
\bibitem [{\citenamefont {Fujikake}\ \emph {et~al.}(2018)\citenamefont
  {Fujikake}, \citenamefont {Deringer}, \citenamefont {Lee}, \citenamefont
  {Krynski}, \citenamefont {Elliott},\ and\ \citenamefont
  {Cs{\'{a}}nyi}}]{Fujikake2018}%
  \BibitemOpen
  \bibfield  {author} {\bibinfo {author} {\bibfnamefont {S.}~\bibnamefont
  {Fujikake}}, \bibinfo {author} {\bibfnamefont {V.~L.}\ \bibnamefont
  {Deringer}}, \bibinfo {author} {\bibfnamefont {T.~H.}\ \bibnamefont {Lee}},
  \bibinfo {author} {\bibfnamefont {M.}~\bibnamefont {Krynski}}, \bibinfo
  {author} {\bibfnamefont {S.~R.}\ \bibnamefont {Elliott}},\ and\ \bibinfo
  {author} {\bibfnamefont {G.}~\bibnamefont {Cs{\'{a}}nyi}},\ }\bibfield
  {title} {\enquote {\bibinfo {title} {{Gaussian approximation potential
  modeling of lithium intercalation in carbon nanostructures}},}\ }\href
  {https://doi.org/10.1063/1.5016317} {\bibfield  {journal} {\bibinfo
  {journal} {J. Chem. Phys.}\ }\textbf {\bibinfo {volume} {148}},\ \bibinfo
  {pages} {241714} (\bibinfo {year} {2018})}\BibitemShut {NoStop}%
\bibitem [{\citenamefont {Unke}\ and\ \citenamefont
  {Meuwly}(2019)}]{unke2019physnet}%
  \BibitemOpen
  \bibfield  {author} {\bibinfo {author} {\bibfnamefont {O.~T.}\ \bibnamefont
  {Unke}}\ and\ \bibinfo {author} {\bibfnamefont {M.}~\bibnamefont {Meuwly}},\
  }\bibfield  {title} {\enquote {\bibinfo {title} {Physnet: A neural network
  for predicting energies, forces, dipole moments, and partial charges},}\
  }\href@noop {} {\bibfield  {journal} {\bibinfo  {journal} {J. Chem. Theory
  Comput.}\ }\textbf {\bibinfo {volume} {15}},\ \bibinfo {pages} {3678--3693}
  (\bibinfo {year} {2019})}\BibitemShut {NoStop}%
\bibitem [{\citenamefont {Cheng}\ \emph
  {et~al.}(2019{\natexlab{a}})\citenamefont {Cheng}, \citenamefont {Welborn},
  \citenamefont {Christensen},\ and\ \citenamefont {{Miller III}}}]{Cheng2019}%
  \BibitemOpen
  \bibfield  {author} {\bibinfo {author} {\bibfnamefont {L.}~\bibnamefont
  {Cheng}}, \bibinfo {author} {\bibfnamefont {M.}~\bibnamefont {Welborn}},
  \bibinfo {author} {\bibfnamefont {A.~S.}\ \bibnamefont {Christensen}},\ and\
  \bibinfo {author} {\bibfnamefont {T.~F.}\ \bibnamefont {{Miller III}}},\
  }\bibfield  {title} {\enquote {\bibinfo {title} {{A universal density matrix
  functional from molecular orbital-based machine learning: Transferability
  across organic molecules}},}\ }\href {https://doi.org/10.1063/1.5088393}
  {\bibfield  {journal} {\bibinfo  {journal} {J. Chem. Phys.}\ }\textbf
  {\bibinfo {volume} {150}},\ \bibinfo {pages} {131103} (\bibinfo {year}
  {2019}{\natexlab{a}})}\BibitemShut {NoStop}%
\bibitem [{\citenamefont {Cheng}\ \emph
  {et~al.}(2019{\natexlab{b}})\citenamefont {Cheng}, \citenamefont {Kovachki},
  \citenamefont {Welborn},\ and\ \citenamefont
  {Miller~III}}]{cheng2019regression}%
  \BibitemOpen
  \bibfield  {author} {\bibinfo {author} {\bibfnamefont {L.}~\bibnamefont
  {Cheng}}, \bibinfo {author} {\bibfnamefont {N.~B.}\ \bibnamefont {Kovachki}},
  \bibinfo {author} {\bibfnamefont {M.}~\bibnamefont {Welborn}},\ and\ \bibinfo
  {author} {\bibfnamefont {T.~F.}\ \bibnamefont {Miller~III}},\ }\bibfield
  {title} {\enquote {\bibinfo {title} {Regression clustering for improved
  accuracy and training costs with molecular-orbital-based machine learning},}\
  }\href@noop {} {\bibfield  {journal} {\bibinfo  {journal} {J. Chem. Theory
  Comput.}\ }\textbf {\bibinfo {volume} {15}},\ \bibinfo {pages} {6668--6677}
  (\bibinfo {year} {2019}{\natexlab{b}})}\BibitemShut {NoStop}%
\bibitem [{\citenamefont {Dick}\ and\ \citenamefont
  {Fernandez-Serra}(2020)}]{dick2020machine}%
  \BibitemOpen
  \bibfield  {author} {\bibinfo {author} {\bibfnamefont {S.}~\bibnamefont
  {Dick}}\ and\ \bibinfo {author} {\bibfnamefont {M.}~\bibnamefont
  {Fernandez-Serra}},\ }\bibfield  {title} {\enquote {\bibinfo {title} {Machine
  learning accurate exchange and correlation functionals of the electronic
  density},}\ }\href@noop {} {\bibfield  {journal} {\bibinfo  {journal} {Nat.
  Commun.}\ }\textbf {\bibinfo {volume} {11}},\ \bibinfo {pages} {1--10}
  (\bibinfo {year} {2020})}\BibitemShut {NoStop}%
\bibitem [{\citenamefont {Chen}\ \emph {et~al.}(2020)\citenamefont {Chen},
  \citenamefont {Zhang}, \citenamefont {Wang},\ and\ \citenamefont
  {E}}]{deephf}%
  \BibitemOpen
  \bibfield  {author} {\bibinfo {author} {\bibfnamefont {Y.}~\bibnamefont
  {Chen}}, \bibinfo {author} {\bibfnamefont {L.}~\bibnamefont {Zhang}},
  \bibinfo {author} {\bibfnamefont {H.}~\bibnamefont {Wang}},\ and\ \bibinfo
  {author} {\bibfnamefont {W.}~\bibnamefont {E}},\ }\bibfield  {title}
  {\enquote {\bibinfo {title} {Ground state energy functional with
  {H}artree--{F}ock efficiency and chemical accuracy},}\ }\href@noop {}
  {\bibfield  {journal} {\bibinfo  {journal} {J. Phys. Chem. A}\ }\textbf
  {\bibinfo {volume} {124}},\ \bibinfo {pages} {7155--7165} (\bibinfo {year}
  {2020})}\BibitemShut {NoStop}%
\bibitem [{\citenamefont {Qiao}\ \emph
  {et~al.}(2020{\natexlab{a}})\citenamefont {Qiao}, \citenamefont {Welborn},
  \citenamefont {Anandkumar}, \citenamefont {Manby},\ and\ \citenamefont
  {Miller~III}}]{qiao2020orbnet}%
  \BibitemOpen
  \bibfield  {author} {\bibinfo {author} {\bibfnamefont {Z.}~\bibnamefont
  {Qiao}}, \bibinfo {author} {\bibfnamefont {M.}~\bibnamefont {Welborn}},
  \bibinfo {author} {\bibfnamefont {A.}~\bibnamefont {Anandkumar}}, \bibinfo
  {author} {\bibfnamefont {F.~R.}\ \bibnamefont {Manby}},\ and\ \bibinfo
  {author} {\bibfnamefont {T.~F.}\ \bibnamefont {Miller~III}},\ }\bibfield
  {title} {\enquote {\bibinfo {title} {Orbnet: Deep learning for quantum
  chemistry using symmetry-adapted atomic-orbital features},}\ }\href@noop {}
  {\bibfield  {journal} {\bibinfo  {journal} {J. Chem. Phys.}\ }\textbf
  {\bibinfo {volume} {153}},\ \bibinfo {pages} {124111} (\bibinfo {year}
  {2020}{\natexlab{a}})}\BibitemShut {NoStop}%
\bibitem [{\citenamefont {Qiao}\ \emph
  {et~al.}(2020{\natexlab{b}})\citenamefont {Qiao}, \citenamefont {Ding},
  \citenamefont {Welborn}, \citenamefont {Bygrave}, \citenamefont {Smith},
  \citenamefont {Anandkumar}, \citenamefont {Manby},\ and\ \citenamefont
  {Miller~III}}]{qiao2020multi}%
  \BibitemOpen
  \bibfield  {author} {\bibinfo {author} {\bibfnamefont {Z.}~\bibnamefont
  {Qiao}}, \bibinfo {author} {\bibfnamefont {F.}~\bibnamefont {Ding}}, \bibinfo
  {author} {\bibfnamefont {M.}~\bibnamefont {Welborn}}, \bibinfo {author}
  {\bibfnamefont {P.~J.}\ \bibnamefont {Bygrave}}, \bibinfo {author}
  {\bibfnamefont {D.~G.}\ \bibnamefont {Smith}}, \bibinfo {author}
  {\bibfnamefont {A.}~\bibnamefont {Anandkumar}}, \bibinfo {author}
  {\bibfnamefont {F.~R.}\ \bibnamefont {Manby}},\ and\ \bibinfo {author}
  {\bibfnamefont {T.~F.}\ \bibnamefont {Miller~III}},\ }\bibfield  {title}
  {\enquote {\bibinfo {title} {Multi-task learning for electronic structure to
  predict and explore molecular potential energy surfaces},}\ }\href@noop {}
  {\bibfield  {journal} {\bibinfo  {journal} {arXiv preprint arXiv:2011.02680}\
  } (\bibinfo {year} {2020}{\natexlab{b}})}\BibitemShut {NoStop}%
\bibitem [{\citenamefont {Hermann}, \citenamefont {Sch{\"a}tzle},\ and\
  \citenamefont {No{\'e}}(2020)}]{hermann2020deep}%
  \BibitemOpen
  \bibfield  {author} {\bibinfo {author} {\bibfnamefont {J.}~\bibnamefont
  {Hermann}}, \bibinfo {author} {\bibfnamefont {Z.}~\bibnamefont
  {Sch{\"a}tzle}},\ and\ \bibinfo {author} {\bibfnamefont {F.}~\bibnamefont
  {No{\'e}}},\ }\bibfield  {title} {\enquote {\bibinfo {title}
  {Deep-neural-network solution of the electronic {S}chr{\"o}dinger
  equation},}\ }\href@noop {} {\bibfield  {journal} {\bibinfo  {journal} {Nat.
  Chem.}\ }\textbf {\bibinfo {volume} {12}},\ \bibinfo {pages} {891--897}
  (\bibinfo {year} {2020})}\BibitemShut {NoStop}%
\bibitem [{\citenamefont {Deringer}\ \emph {et~al.}(2021)\citenamefont
  {Deringer}, \citenamefont {Bernstein}, \citenamefont {Cs{\'a}nyi},
  \citenamefont {Ben~Mahmoud}, \citenamefont {Ceriotti}, \citenamefont
  {Wilson}, \citenamefont {Drabold},\ and\ \citenamefont
  {Elliott}}]{deringer2021origins}%
  \BibitemOpen
  \bibfield  {author} {\bibinfo {author} {\bibfnamefont {V.~L.}\ \bibnamefont
  {Deringer}}, \bibinfo {author} {\bibfnamefont {N.}~\bibnamefont {Bernstein}},
  \bibinfo {author} {\bibfnamefont {G.}~\bibnamefont {Cs{\'a}nyi}}, \bibinfo
  {author} {\bibfnamefont {C.}~\bibnamefont {Ben~Mahmoud}}, \bibinfo {author}
  {\bibfnamefont {M.}~\bibnamefont {Ceriotti}}, \bibinfo {author}
  {\bibfnamefont {M.}~\bibnamefont {Wilson}}, \bibinfo {author} {\bibfnamefont
  {D.~A.}\ \bibnamefont {Drabold}},\ and\ \bibinfo {author} {\bibfnamefont
  {S.~R.}\ \bibnamefont {Elliott}},\ }\bibfield  {title} {\enquote {\bibinfo
  {title} {Origins of structural and electronic transitions in disordered
  silicon},}\ }\href@noop {} {\bibfield  {journal} {\bibinfo  {journal}
  {Nature}\ }\textbf {\bibinfo {volume} {589}},\ \bibinfo {pages} {59--64}
  (\bibinfo {year} {2021})}\BibitemShut {NoStop}%
\bibitem [{\citenamefont {Christensen}\ \emph {et~al.}(2021)\citenamefont
  {Christensen}, \citenamefont {Sirumalla}, \citenamefont {Qiao}, \citenamefont
  {O’Connor}, \citenamefont {Smith}, \citenamefont {Ding}, \citenamefont
  {Bygrave}, \citenamefont {Anandkumar}, \citenamefont {Welborn}, \citenamefont
  {Manby} \emph {et~al.}}]{christensen2021orbnet}%
  \BibitemOpen
  \bibfield  {author} {\bibinfo {author} {\bibfnamefont {A.~S.}\ \bibnamefont
  {Christensen}}, \bibinfo {author} {\bibfnamefont {S.~K.}\ \bibnamefont
  {Sirumalla}}, \bibinfo {author} {\bibfnamefont {Z.}~\bibnamefont {Qiao}},
  \bibinfo {author} {\bibfnamefont {M.~B.}\ \bibnamefont {O’Connor}},
  \bibinfo {author} {\bibfnamefont {D.~G.}\ \bibnamefont {Smith}}, \bibinfo
  {author} {\bibfnamefont {F.}~\bibnamefont {Ding}}, \bibinfo {author}
  {\bibfnamefont {P.~J.}\ \bibnamefont {Bygrave}}, \bibinfo {author}
  {\bibfnamefont {A.}~\bibnamefont {Anandkumar}}, \bibinfo {author}
  {\bibfnamefont {M.}~\bibnamefont {Welborn}}, \bibinfo {author} {\bibfnamefont
  {F.~R.}\ \bibnamefont {Manby}}, \emph {et~al.},\ }\bibfield  {title}
  {\enquote {\bibinfo {title} {Orbnet denali: A machine learning potential for
  biological and organic chemistry with semi-empirical cost and dft
  accuracy},}\ }\href@noop {} {\bibfield  {journal} {\bibinfo  {journal} {J.
  Chem. Phys.}\ }\textbf {\bibinfo {volume} {155}},\ \bibinfo {pages} {204103}
  (\bibinfo {year} {2021})}\BibitemShut {NoStop}%
\bibitem [{\citenamefont {Husch}\ \emph {et~al.}(2021)\citenamefont {Husch},
  \citenamefont {Sun}, \citenamefont {Cheng}, \citenamefont {Lee},\ and\
  \citenamefont {Miller~III}}]{husch2020improved}%
  \BibitemOpen
  \bibfield  {author} {\bibinfo {author} {\bibfnamefont {T.}~\bibnamefont
  {Husch}}, \bibinfo {author} {\bibfnamefont {J.}~\bibnamefont {Sun}}, \bibinfo
  {author} {\bibfnamefont {L.}~\bibnamefont {Cheng}}, \bibinfo {author}
  {\bibfnamefont {S.~J.}\ \bibnamefont {Lee}},\ and\ \bibinfo {author}
  {\bibfnamefont {T.~F.}\ \bibnamefont {Miller~III}},\ }\bibfield  {title}
  {\enquote {\bibinfo {title} {Improved accuracy and transferability of
  molecular-orbital-based machine learning: Organics, transition-metal
  complexes, non-covalent interactions, and transition states},}\ }\href@noop
  {} {\bibfield  {journal} {\bibinfo  {journal} {J. Chem. Phys.}\ }\textbf
  {\bibinfo {volume} {154}},\ \bibinfo {pages} {064108} (\bibinfo {year}
  {2021})}\BibitemShut {NoStop}%
\bibitem [{\citenamefont {Lee}\ \emph {et~al.}(2021)\citenamefont {Lee},
  \citenamefont {Husch}, \citenamefont {Ding},\ and\ \citenamefont
  {Miller~III}}]{lee2020analytical}%
  \BibitemOpen
  \bibfield  {author} {\bibinfo {author} {\bibfnamefont {S.~J.~R.}\
  \bibnamefont {Lee}}, \bibinfo {author} {\bibfnamefont {T.}~\bibnamefont
  {Husch}}, \bibinfo {author} {\bibfnamefont {F.}~\bibnamefont {Ding}},\ and\
  \bibinfo {author} {\bibfnamefont {T.~F.}\ \bibnamefont {Miller~III}},\
  }\bibfield  {title} {\enquote {\bibinfo {title} {Analytical gradients for
  molecular-orbital-based machine learning},}\ }\href
  {https://doi.org/10.1063/5.0040782} {\bibfield  {journal} {\bibinfo
  {journal} {J. Chem. Phys.}\ }\textbf {\bibinfo {volume} {154}},\ \bibinfo
  {pages} {124120} (\bibinfo {year} {2021})}\BibitemShut {NoStop}%
\bibitem [{\citenamefont {Sun}, \citenamefont {Cheng},\ and\ \citenamefont
  {Miller~III}(2021)}]{sun2021molecular}%
  \BibitemOpen
  \bibfield  {author} {\bibinfo {author} {\bibfnamefont {J.}~\bibnamefont
  {Sun}}, \bibinfo {author} {\bibfnamefont {L.}~\bibnamefont {Cheng}},\ and\
  \bibinfo {author} {\bibfnamefont {T.~F.}\ \bibnamefont {Miller~III}},\
  }\bibfield  {title} {\enquote {\bibinfo {title} {Molecular energy learning
  using alternative blackbox matrix-matrix multiplication algorithm for exact
  gaussian process},}\ }in\ \href {https://openreview.net/forum?id=lyJ9BRKUzms}
  {\emph {\bibinfo {booktitle} {NeurIPS 2021 AI for Science Workshop}}}\
  (\bibinfo {year} {2021})\BibitemShut {NoStop}%
\bibitem [{\citenamefont {Karandashev}\ and\ \citenamefont {von
  Lilienfeld}(2022)}]{Karandashev2022}%
  \BibitemOpen
  \bibfield  {author} {\bibinfo {author} {\bibfnamefont {K.}~\bibnamefont
  {Karandashev}}\ and\ \bibinfo {author} {\bibfnamefont {O.~A.}\ \bibnamefont
  {von Lilienfeld}},\ }\bibfield  {title} {\enquote {\bibinfo {title} {An
  orbital-based representation for accurate quantum machine learning},}\ }\href
  {https://doi.org/10.1063/5.0083301} {\bibfield  {journal} {\bibinfo
  {journal} {J. Chem. Phys.}\ }\textbf {\bibinfo {volume} {156}},\ \bibinfo
  {pages} {114101} (\bibinfo {year} {2022})}\BibitemShut {NoStop}%
\bibitem [{\citenamefont {Veit}\ \emph {et~al.}(2020)\citenamefont {Veit},
  \citenamefont {Wilkins}, \citenamefont {Yang}, \citenamefont {DiStasio},\
  and\ \citenamefont {Ceriotti}}]{veit2020predicting}%
  \BibitemOpen
  \bibfield  {author} {\bibinfo {author} {\bibfnamefont {M.}~\bibnamefont
  {Veit}}, \bibinfo {author} {\bibfnamefont {D.~M.}\ \bibnamefont {Wilkins}},
  \bibinfo {author} {\bibfnamefont {Y.}~\bibnamefont {Yang}}, \bibinfo {author}
  {\bibfnamefont {R.~A.}\ \bibnamefont {DiStasio}},\ and\ \bibinfo {author}
  {\bibfnamefont {M.}~\bibnamefont {Ceriotti}},\ }\bibfield  {title} {\enquote
  {\bibinfo {title} {Predicting molecular dipole moments by combining atomic
  partial charges and atomic dipoles},}\ }\href
  {https://doi.org/10.1063/5.0009106} {\bibfield  {journal} {\bibinfo
  {journal} {J. Chem. Phys.}\ }\textbf {\bibinfo {volume} {153}},\ \bibinfo
  {pages} {024113} (\bibinfo {year} {2020})}\BibitemShut {NoStop}%
\bibitem [{\citenamefont {Qiao}\ \emph {et~al.}(2021)\citenamefont {Qiao},
  \citenamefont {Christensen}, \citenamefont {Welborn}, \citenamefont {Manby},
  \citenamefont {Anandkumar},\ and\ \citenamefont {Miller}}]{orbnetequi}%
  \BibitemOpen
  \bibfield  {author} {\bibinfo {author} {\bibfnamefont {Z.}~\bibnamefont
  {Qiao}}, \bibinfo {author} {\bibfnamefont {A.~S.}\ \bibnamefont
  {Christensen}}, \bibinfo {author} {\bibfnamefont {M.}~\bibnamefont
  {Welborn}}, \bibinfo {author} {\bibfnamefont {F.~R.}\ \bibnamefont {Manby}},
  \bibinfo {author} {\bibfnamefont {A.}~\bibnamefont {Anandkumar}},\ and\
  \bibinfo {author} {\bibfnamefont {T.~F.}\ \bibnamefont {Miller}},\ }\bibfield
   {title} {\enquote {\bibinfo {title} {Informing geometric deep learning with
  electronic interactions to accelerate quantum chemistry},}\ }\href@noop {}
  {\bibfield  {journal} {\bibinfo  {journal} {arXiv preprint arXiv:2011.02680}\
  } (\bibinfo {year} {2021})}\BibitemShut {NoStop}%
\bibitem [{\citenamefont {Klicpera}\ \emph {et~al.}(2020)\citenamefont
  {Klicpera}, \citenamefont {Giri}, \citenamefont {Margraf},\ and\
  \citenamefont {G{\"u}nnemann}}]{klicpera2020fast}%
  \BibitemOpen
  \bibfield  {author} {\bibinfo {author} {\bibfnamefont {J.}~\bibnamefont
  {Klicpera}}, \bibinfo {author} {\bibfnamefont {S.}~\bibnamefont {Giri}},
  \bibinfo {author} {\bibfnamefont {J.~T.}\ \bibnamefont {Margraf}},\ and\
  \bibinfo {author} {\bibfnamefont {S.}~\bibnamefont {G{\"u}nnemann}},\
  }\bibfield  {title} {\enquote {\bibinfo {title} {Fast and uncertainty-aware
  directional message passing for non-equilibrium molecules},}\ }\href@noop {}
  {\bibfield  {journal} {\bibinfo  {journal} {arXiv preprint arXiv:2011.14115}\
  } (\bibinfo {year} {2020})}\BibitemShut {NoStop}%
\bibitem [{\citenamefont {Liu}\ \emph {et~al.}(2022)\citenamefont {Liu},
  \citenamefont {Wang}, \citenamefont {Liu}, \citenamefont {Lin}, \citenamefont
  {Zhang}, \citenamefont {Oztekin},\ and\ \citenamefont
  {Ji}}]{liu2022spherical}%
  \BibitemOpen
  \bibfield  {author} {\bibinfo {author} {\bibfnamefont {Y.}~\bibnamefont
  {Liu}}, \bibinfo {author} {\bibfnamefont {L.}~\bibnamefont {Wang}}, \bibinfo
  {author} {\bibfnamefont {M.}~\bibnamefont {Liu}}, \bibinfo {author}
  {\bibfnamefont {Y.}~\bibnamefont {Lin}}, \bibinfo {author} {\bibfnamefont
  {X.}~\bibnamefont {Zhang}}, \bibinfo {author} {\bibfnamefont
  {B.}~\bibnamefont {Oztekin}},\ and\ \bibinfo {author} {\bibfnamefont
  {S.}~\bibnamefont {Ji}},\ }\bibfield  {title} {\enquote {\bibinfo {title}
  {Spherical message passing for 3d molecular graphs},}\ }in\ \href
  {https://openreview.net/forum?id=givsRXsOt9r} {\emph {\bibinfo {booktitle}
  {International Conference on Learning Representations}}}\ (\bibinfo {year}
  {2022})\BibitemShut {NoStop}%
\bibitem [{\citenamefont {Sch{\"u}tt}, \citenamefont {Unke},\ and\
  \citenamefont {Gastegger}(2021)}]{painn}%
  \BibitemOpen
  \bibfield  {author} {\bibinfo {author} {\bibfnamefont {K.}~\bibnamefont
  {Sch{\"u}tt}}, \bibinfo {author} {\bibfnamefont {O.}~\bibnamefont {Unke}},\
  and\ \bibinfo {author} {\bibfnamefont {M.}~\bibnamefont {Gastegger}},\
  }\bibfield  {title} {\enquote {\bibinfo {title} {Equivariant message passing
  for the prediction of tensorial properties and molecular spectra},}\ }in\
  \href {https://proceedings.mlr.press/v139/schutt21a.html} {\emph {\bibinfo
  {booktitle} {Proceedings of the 38th International Conference on Machine
  Learning}}},\ \bibinfo {series} {Proceedings of Machine Learning Research},
  Vol.\ \bibinfo {volume} {139},\ \bibinfo {editor} {edited by\ \bibinfo
  {editor} {\bibfnamefont {M.}~\bibnamefont {Meila}}\ and\ \bibinfo {editor}
  {\bibfnamefont {T.}~\bibnamefont {Zhang}}}\ (\bibinfo  {publisher} {PMLR},\
  \bibinfo {year} {2021})\ pp.\ \bibinfo {pages} {9377--9388}\BibitemShut
  {NoStop}%
\bibitem [{\citenamefont {Faber}\ \emph {et~al.}(2018)\citenamefont {Faber},
  \citenamefont {Christensen}, \citenamefont {Huang},\ and\ \citenamefont
  {Von~Lilienfeld}}]{faber2018alchemical}%
  \BibitemOpen
  \bibfield  {author} {\bibinfo {author} {\bibfnamefont {F.~A.}\ \bibnamefont
  {Faber}}, \bibinfo {author} {\bibfnamefont {A.~S.}\ \bibnamefont
  {Christensen}}, \bibinfo {author} {\bibfnamefont {B.}~\bibnamefont {Huang}},\
  and\ \bibinfo {author} {\bibfnamefont {O.~A.}\ \bibnamefont
  {Von~Lilienfeld}},\ }\bibfield  {title} {\enquote {\bibinfo {title}
  {Alchemical and structural distribution based representation for universal
  quantum machine learning},}\ }\href@noop {} {\bibfield  {journal} {\bibinfo
  {journal} {J. Chem. Phys.}\ }\textbf {\bibinfo {volume} {148}},\ \bibinfo
  {pages} {241717} (\bibinfo {year} {2018})}\BibitemShut {NoStop}%
\bibitem [{\citenamefont {Huang}\ and\ \citenamefont {von
  Lilienfeld}(2020)}]{huang2020quantum}%
  \BibitemOpen
  \bibfield  {author} {\bibinfo {author} {\bibfnamefont {B.}~\bibnamefont
  {Huang}}\ and\ \bibinfo {author} {\bibfnamefont {O.~A.}\ \bibnamefont {von
  Lilienfeld}},\ }\bibfield  {title} {\enquote {\bibinfo {title} {Quantum
  machine learning using atom-in-molecule-based fragments selected on the
  fly},}\ }\href@noop {} {\bibfield  {journal} {\bibinfo  {journal} {Nat.
  Chem.}\ }\textbf {\bibinfo {volume} {12}},\ \bibinfo {pages} {945--951}
  (\bibinfo {year} {2020})}\BibitemShut {NoStop}%
\bibitem [{\citenamefont {Bart{\'o}k}\ \emph {et~al.}(2017)\citenamefont
  {Bart{\'o}k}, \citenamefont {De}, \citenamefont {Poelking}, \citenamefont
  {Bernstein}, \citenamefont {Kermode}, \citenamefont {Cs{\'a}nyi},\ and\
  \citenamefont {Ceriotti}}]{bartok2017machine}%
  \BibitemOpen
  \bibfield  {author} {\bibinfo {author} {\bibfnamefont {A.~P.}\ \bibnamefont
  {Bart{\'o}k}}, \bibinfo {author} {\bibfnamefont {S.}~\bibnamefont {De}},
  \bibinfo {author} {\bibfnamefont {C.}~\bibnamefont {Poelking}}, \bibinfo
  {author} {\bibfnamefont {N.}~\bibnamefont {Bernstein}}, \bibinfo {author}
  {\bibfnamefont {J.~R.}\ \bibnamefont {Kermode}}, \bibinfo {author}
  {\bibfnamefont {G.}~\bibnamefont {Cs{\'a}nyi}},\ and\ \bibinfo {author}
  {\bibfnamefont {M.}~\bibnamefont {Ceriotti}},\ }\bibfield  {title} {\enquote
  {\bibinfo {title} {Machine learning unifies the modeling of materials and
  molecules},}\ }\href@noop {} {\bibfield  {journal} {\bibinfo  {journal} {Sci.
  Adv.}\ }\textbf {\bibinfo {volume} {3}},\ \bibinfo {pages} {e1701816}
  (\bibinfo {year} {2017})}\BibitemShut {NoStop}%
\bibitem [{\citenamefont {Christensen}, \citenamefont {Faber},\ and\
  \citenamefont {von Lilienfeld}(2019)}]{OQML}%
  \BibitemOpen
  \bibfield  {author} {\bibinfo {author} {\bibfnamefont {A.~S.}\ \bibnamefont
  {Christensen}}, \bibinfo {author} {\bibfnamefont {F.~A.}\ \bibnamefont
  {Faber}},\ and\ \bibinfo {author} {\bibfnamefont {O.~A.}\ \bibnamefont {von
  Lilienfeld}},\ }\bibfield  {title} {\enquote {\bibinfo {title} {Operators in
  quantum machine learning: Response properties in chemical space},}\ }\href
  {https://doi.org/10.1063/1.5053562} {\bibfield  {journal} {\bibinfo
  {journal} {J. Chem. Phys.}\ }\textbf {\bibinfo {volume} {150}},\ \bibinfo
  {pages} {064105} (\bibinfo {year} {2019})}\BibitemShut {NoStop}%
\bibitem [{\citenamefont {Cheng}, \citenamefont {Sun},\ and\ \citenamefont
  {Miller~III}(2022{\natexlab{a}})}]{cheng2022}%
  \BibitemOpen
  \bibfield  {author} {\bibinfo {author} {\bibfnamefont {L.}~\bibnamefont
  {Cheng}}, \bibinfo {author} {\bibfnamefont {J.}~\bibnamefont {Sun}},\ and\
  \bibinfo {author} {\bibfnamefont {T.~F.}\ \bibnamefont {Miller~III}},\
  }\bibfield  {title} {\enquote {\bibinfo {title} {Improved accuracy of
  molecular energy learning via unsupervised clustering for organic chemical
  space with molecular-orbital-based machine learning},}\ }\href@noop {}
  {\bibfield  {journal} {\bibinfo  {journal} {arXiv preprint arXiv:2204.09831}\
  } (\bibinfo {year} {2022}{\natexlab{a}})}\BibitemShut {NoStop}%
\bibitem [{\citenamefont {Gonze}(1995)}]{gonze1995adiabatic}%
  \BibitemOpen
  \bibfield  {author} {\bibinfo {author} {\bibfnamefont {X.}~\bibnamefont
  {Gonze}},\ }\bibfield  {title} {\enquote {\bibinfo {title} {Adiabatic
  density-functional perturbation theory},}\ }\href@noop {} {\bibfield
  {journal} {\bibinfo  {journal} {Phys. Rev. A}\ }\textbf {\bibinfo {volume}
  {52}},\ \bibinfo {pages} {1096} (\bibinfo {year} {1995})}\BibitemShut
  {NoStop}%
\bibitem [{\citenamefont {Casida}(2009)}]{LRTDDFT}%
  \BibitemOpen
  \bibfield  {author} {\bibinfo {author} {\bibfnamefont {M.~E.}\ \bibnamefont
  {Casida}},\ }\bibfield  {title} {\enquote {\bibinfo {title} {Time-dependent
  density-functional theory for molecules and molecular solids},}\ }\href@noop
  {} {\bibfield  {journal} {\bibinfo  {journal} {J. Mol. Struct.: THEOCHEM}\
  }\textbf {\bibinfo {volume} {914}},\ \bibinfo {pages} {3--18} (\bibinfo
  {year} {2009})}\BibitemShut {NoStop}%
\bibitem [{\citenamefont {Grunenberg}(2011)}]{grunenberg2011computational}%
  \BibitemOpen
  \bibfield  {author} {\bibinfo {author} {\bibfnamefont {J.}~\bibnamefont
  {Grunenberg}},\ }\href@noop {} {\emph {\bibinfo {title} {Computational
  spectroscopy: methods, experiments and \\applications}}}\ (\bibinfo
  {publisher} {John Wiley \& Sons},\ \bibinfo {year} {2011})\BibitemShut
  {NoStop}%
\bibitem [{\citenamefont {Hait}\ and\ \citenamefont
  {Head-Gordon}(2018)}]{hait2018accurate}%
  \BibitemOpen
  \bibfield  {author} {\bibinfo {author} {\bibfnamefont {D.}~\bibnamefont
  {Hait}}\ and\ \bibinfo {author} {\bibfnamefont {M.}~\bibnamefont
  {Head-Gordon}},\ }\bibfield  {title} {\enquote {\bibinfo {title} {How
  accurate is density functional theory at predicting dipole moments? {A}n
  assessment using a new database of 200 benchmark values},}\ }\href@noop {}
  {\bibfield  {journal} {\bibinfo  {journal} {J. Chem. Theory Comput.}\
  }\textbf {\bibinfo {volume} {14}},\ \bibinfo {pages} {1969--1981} (\bibinfo
  {year} {2018})}\BibitemShut {NoStop}%
\bibitem [{\citenamefont {Darley}, \citenamefont {Handley},\ and\ \citenamefont
  {Popelier}(2008)}]{darley2008beyond}%
  \BibitemOpen
  \bibfield  {author} {\bibinfo {author} {\bibfnamefont {M.~G.}\ \bibnamefont
  {Darley}}, \bibinfo {author} {\bibfnamefont {C.~M.}\ \bibnamefont
  {Handley}},\ and\ \bibinfo {author} {\bibfnamefont {P.~L.}\ \bibnamefont
  {Popelier}},\ }\bibfield  {title} {\enquote {\bibinfo {title} {Beyond point
  charges: dynamic polarization from neural net predicted multipole moments},}\
  }\href@noop {} {\bibfield  {journal} {\bibinfo  {journal} {J. Chem. Theory
  Comput.}\ }\textbf {\bibinfo {volume} {4}},\ \bibinfo {pages} {1435--1448}
  (\bibinfo {year} {2008})}\BibitemShut {NoStop}%
\bibitem [{\citenamefont {Wilkins}\ \emph {et~al.}(2019)\citenamefont
  {Wilkins}, \citenamefont {Grisafi}, \citenamefont {Yang}, \citenamefont
  {Lao}, \citenamefont {DiStasio},\ and\ \citenamefont
  {Ceriotti}}]{wilkins2019accurate}%
  \BibitemOpen
  \bibfield  {author} {\bibinfo {author} {\bibfnamefont {D.~M.}\ \bibnamefont
  {Wilkins}}, \bibinfo {author} {\bibfnamefont {A.}~\bibnamefont {Grisafi}},
  \bibinfo {author} {\bibfnamefont {Y.}~\bibnamefont {Yang}}, \bibinfo {author}
  {\bibfnamefont {K.~U.}\ \bibnamefont {Lao}}, \bibinfo {author} {\bibfnamefont
  {R.~A.}\ \bibnamefont {DiStasio}},\ and\ \bibinfo {author} {\bibfnamefont
  {M.}~\bibnamefont {Ceriotti}},\ }\bibfield  {title} {\enquote {\bibinfo
  {title} {Accurate molecular polarizabilities with coupled cluster theory and
  machine learning},}\ }\href@noop {} {\bibfield  {journal} {\bibinfo
  {journal} {Proc. Natl. Acad. Sci. U.S.A.}\ }\textbf {\bibinfo {volume}
  {116}},\ \bibinfo {pages} {3401--3406} (\bibinfo {year} {2019})}\BibitemShut
  {NoStop}%
\bibitem [{\citenamefont {Rasmussen}\ and\ \citenamefont
  {Williams}(2006)}]{GPbook}%
  \BibitemOpen
  \bibfield  {author} {\bibinfo {author} {\bibfnamefont {C.~E.}\ \bibnamefont
  {Rasmussen}}\ and\ \bibinfo {author} {\bibfnamefont {C.~K.~I.}\ \bibnamefont
  {Williams}},\ }\href {http://www.gaussianprocess.org/gpml/chapters/RW.pdf}
  {\emph {\bibinfo {title} {Gaussian processes for machine learning}}}\
  (\bibinfo  {publisher} {MIT Press},\ \bibinfo {address} {Cambridge, MA},\
  \bibinfo {year} {2006})\BibitemShut {NoStop}%
\bibitem [{\citenamefont {Cheng}, \citenamefont {Sun},\ and\ \citenamefont
  {Miller~III}(2022{\natexlab{b}})}]{cheng2022accurate}%
  \BibitemOpen
  \bibfield  {author} {\bibinfo {author} {\bibfnamefont {L.}~\bibnamefont
  {Cheng}}, \bibinfo {author} {\bibfnamefont {J.}~\bibnamefont {Sun}},\ and\
  \bibinfo {author} {\bibfnamefont {T.~F.}\ \bibnamefont {Miller~III}},\
  }\bibfield  {title} {\enquote {\bibinfo {title} {Accurate
  molecular-orbital-based machine learning energies via unsupervised clustering
  of chemical space},}\ }\href@noop {} {\bibfield  {journal} {\bibinfo
  {journal} {arXiv preprint arXiv:2204.09831}\ } (\bibinfo {year}
  {2022}{\natexlab{b}})}\BibitemShut {NoStop}%
\bibitem [{\citenamefont {Ramakrishnan}\ \emph {et~al.}(2014)\citenamefont
  {Ramakrishnan}, \citenamefont {Dral}, \citenamefont {Rupp},\ and\
  \citenamefont {Von~Lilienfeld}}]{ramakrishnan2014quantum}%
  \BibitemOpen
  \bibfield  {author} {\bibinfo {author} {\bibfnamefont {R.}~\bibnamefont
  {Ramakrishnan}}, \bibinfo {author} {\bibfnamefont {P.~O.}\ \bibnamefont
  {Dral}}, \bibinfo {author} {\bibfnamefont {M.}~\bibnamefont {Rupp}},\ and\
  \bibinfo {author} {\bibfnamefont {O.~A.}\ \bibnamefont {Von~Lilienfeld}},\
  }\bibfield  {title} {\enquote {\bibinfo {title} {Quantum chemistry structures
  and properties of 134 kilo molecules},}\ }\href@noop {} {\bibfield  {journal}
  {\bibinfo  {journal} {Sci. Data}\ }\textbf {\bibinfo {volume} {1}},\ \bibinfo
  {pages} {1--7} (\bibinfo {year} {2014})}\BibitemShut {NoStop}%
\bibitem [{\citenamefont {Nesbet}(1958)}]{Nesbet1958}%
  \BibitemOpen
  \bibfield  {author} {\bibinfo {author} {\bibfnamefont {R.~K.}\ \bibnamefont
  {Nesbet}},\ }\bibfield  {title} {\enquote {\bibinfo {title} {{Brueckner's
  theory and the method of superposition of configurations}},}\ }\href@noop {}
  {\bibfield  {journal} {\bibinfo  {journal} {Phys. Rev.}\ }\textbf {\bibinfo
  {volume} {109}},\ \bibinfo {pages} {1632} (\bibinfo {year}
  {1958})}\BibitemShut {NoStop}%
\bibitem [{\citenamefont {Saebo}\ and\ \citenamefont {Pulay}(1993)}]{LMP2}%
  \BibitemOpen
  \bibfield  {author} {\bibinfo {author} {\bibfnamefont {S.}~\bibnamefont
  {Saebo}}\ and\ \bibinfo {author} {\bibfnamefont {P.}~\bibnamefont {Pulay}},\
  }\bibfield  {title} {\enquote {\bibinfo {title} {{Local treatment of electron
  correlation}},}\ }\href {https://doi.org/10.1146/annurev.pc.44.100193.001241}
  {\bibfield  {journal} {\bibinfo  {journal} {Annu. Rev. Phys. Chem.}\ }\textbf
  {\bibinfo {volume} {44}},\ \bibinfo {pages} {213--236} (\bibinfo {year}
  {1993})}\BibitemShut {NoStop}%
\bibitem [{\citenamefont {Hampel}\ and\ \citenamefont {Werner}(1996)}]{LCCSD}%
  \BibitemOpen
  \bibfield  {author} {\bibinfo {author} {\bibfnamefont {C.}~\bibnamefont
  {Hampel}}\ and\ \bibinfo {author} {\bibfnamefont {H.~J.}\ \bibnamefont
  {Werner}},\ }\bibfield  {title} {\enquote {\bibinfo {title} {{Local treatment
  of electron correlation in coupled cluster theory}},}\ }\href
  {https://doi.org/10.1063/1.471289} {\bibfield  {journal} {\bibinfo  {journal}
  {J. Chem. Phys.}\ }\textbf {\bibinfo {volume} {104}},\ \bibinfo {pages}
  {6286--6297} (\bibinfo {year} {1996})}\BibitemShut {NoStop}%
\bibitem [{\citenamefont {Sch{\"{u}}tz}(2000)}]{LCCSDT}%
  \BibitemOpen
  \bibfield  {author} {\bibinfo {author} {\bibfnamefont {M.}~\bibnamefont
  {Sch{\"{u}}tz}},\ }\bibfield  {title} {\enquote {\bibinfo {title} {{Low-order
  scaling local electron correlation methods. III. Linear scaling local
  perturbative triples correction (T)}},}\ }\href
  {https://doi.org/10.1063/1.1323265} {\bibfield  {journal} {\bibinfo
  {journal} {J. Chem. Phys.}\ }\textbf {\bibinfo {volume} {113}},\ \bibinfo
  {pages} {9986--10001} (\bibinfo {year} {2000})}\BibitemShut {NoStop}%
\bibitem [{\citenamefont {Parzen}(1999)}]{parzen1999stochastic}%
  \BibitemOpen
  \bibfield  {author} {\bibinfo {author} {\bibfnamefont {E.}~\bibnamefont
  {Parzen}},\ }\href@noop {} {\emph {\bibinfo {title} {Stochastic processes}}}\
  (\bibinfo  {publisher} {SIAM},\ \bibinfo {year} {1999})\BibitemShut {NoStop}%
\bibitem [{\citenamefont {Marsden}\ and\ \citenamefont
  {Tromba}(2003)}]{marsden2003vector}%
  \BibitemOpen
  \bibfield  {author} {\bibinfo {author} {\bibfnamefont {J.~E.}\ \bibnamefont
  {Marsden}}\ and\ \bibinfo {author} {\bibfnamefont {A.}~\bibnamefont
  {Tromba}},\ }\href@noop {} {\emph {\bibinfo {title} {Vector calculus}}}\
  (\bibinfo  {publisher} {Macmillan},\ \bibinfo {year} {2003})\BibitemShut
  {NoStop}%
\bibitem [{\citenamefont {Chmiela}\ \emph {et~al.}(2017)\citenamefont
  {Chmiela}, \citenamefont {Tkatchenko}, \citenamefont {Sauceda}, \citenamefont
  {Poltavsky}, \citenamefont {Sch{\"u}tt},\ and\ \citenamefont
  {M{\"u}ller}}]{GDML}%
  \BibitemOpen
  \bibfield  {author} {\bibinfo {author} {\bibfnamefont {S.}~\bibnamefont
  {Chmiela}}, \bibinfo {author} {\bibfnamefont {A.}~\bibnamefont {Tkatchenko}},
  \bibinfo {author} {\bibfnamefont {H.~E.}\ \bibnamefont {Sauceda}}, \bibinfo
  {author} {\bibfnamefont {I.}~\bibnamefont {Poltavsky}}, \bibinfo {author}
  {\bibfnamefont {K.~T.}\ \bibnamefont {Sch{\"u}tt}},\ and\ \bibinfo {author}
  {\bibfnamefont {K.-R.}\ \bibnamefont {M{\"u}ller}},\ }\bibfield  {title}
  {\enquote {\bibinfo {title} {Machine learning of accurate energy-conserving
  molecular force fields},}\ }\href@noop {} {\bibfield  {journal} {\bibinfo
  {journal} {Sci. Adv.}\ }\textbf {\bibinfo {volume} {3}},\ \bibinfo {pages}
  {e1603015} (\bibinfo {year} {2017})}\BibitemShut {NoStop}%
\bibitem [{\citenamefont {Manby}\ \emph {et~al.}(2019)\citenamefont {Manby},
  \citenamefont {Miller~III}, \citenamefont {Bygrave}, \citenamefont {Ding},
  \citenamefont {Dresselhaus}, \citenamefont {Batista-Romero}, \citenamefont
  {Buccheri}, \citenamefont {Bungey}, \citenamefont {Lee}, \citenamefont
  {Meli}, \citenamefont {Miyamoto}, \citenamefont {Steinmann}, \citenamefont
  {Tsuchiya}, \citenamefont {Welborn}, \citenamefont {Wiles},\ and\
  \citenamefont {Williams}}]{manby2019entos}%
  \BibitemOpen
  \bibfield  {author} {\bibinfo {author} {\bibfnamefont {F.~R.}\ \bibnamefont
  {Manby}}, \bibinfo {author} {\bibfnamefont {T.~F.}\ \bibnamefont
  {Miller~III}}, \bibinfo {author} {\bibfnamefont {P.}~\bibnamefont {Bygrave}},
  \bibinfo {author} {\bibfnamefont {F.}~\bibnamefont {Ding}}, \bibinfo {author}
  {\bibfnamefont {T.}~\bibnamefont {Dresselhaus}}, \bibinfo {author}
  {\bibfnamefont {F.}~\bibnamefont {Batista-Romero}}, \bibinfo {author}
  {\bibfnamefont {A.}~\bibnamefont {Buccheri}}, \bibinfo {author}
  {\bibfnamefont {C.}~\bibnamefont {Bungey}}, \bibinfo {author} {\bibfnamefont
  {S.~J.~R.}\ \bibnamefont {Lee}}, \bibinfo {author} {\bibfnamefont
  {R.}~\bibnamefont {Meli}}, \bibinfo {author} {\bibfnamefont {K.}~\bibnamefont
  {Miyamoto}}, \bibinfo {author} {\bibfnamefont {C.}~\bibnamefont {Steinmann}},
  \bibinfo {author} {\bibfnamefont {T.}~\bibnamefont {Tsuchiya}}, \bibinfo
  {author} {\bibfnamefont {M.}~\bibnamefont {Welborn}}, \bibinfo {author}
  {\bibfnamefont {T.}~\bibnamefont {Wiles}},\ and\ \bibinfo {author}
  {\bibfnamefont {Z.}~\bibnamefont {Williams}},\ }\bibfield  {title} {\enquote
  {\bibinfo {title} {entos: {A} {Quantum} {Molecular} {Simulation}
  {Package}},}\ }\href {https://doi.org/10.26434/chemrxiv.7762646.v2}
  {\bibfield  {journal} {\bibinfo  {journal} {ChemRxiv}\ } (\bibinfo {year}
  {2019}),\ 10.26434/chemrxiv.7762646.v2}\BibitemShut {NoStop}%
\bibitem [{\citenamefont {Dunning}(1989)}]{Dunning1989}%
  \BibitemOpen
  \bibfield  {author} {\bibinfo {author} {\bibfnamefont {T.~H.}\ \bibnamefont
  {Dunning}},\ }\bibfield  {title} {\enquote {\bibinfo {title} {{Gaussian basis
  sets for use in correlated molecular calculations. I. The atoms boron through
  neon and hydrogen}},}\ }\href {https://doi.org/10.1063/1.456153} {\bibfield
  {journal} {\bibinfo  {journal} {J. Chem. Phys.}\ }\textbf {\bibinfo {volume}
  {90}},\ \bibinfo {pages} {1007} (\bibinfo {year} {1989})}\BibitemShut
  {NoStop}%
\bibitem [{\citenamefont {Weigend}(2002)}]{weigend_fully_2002}%
  \BibitemOpen
  \bibfield  {author} {\bibinfo {author} {\bibfnamefont {F.}~\bibnamefont
  {Weigend}},\ }\bibfield  {title} {\enquote {\bibinfo {title} {A fully direct
  {RI}-{HF} algorithm: {Implementation}, optimised auxiliary basis sets,
  demonstration of accuracy and efficiency},}\ }\href@noop {} {\bibfield
  {journal} {\bibinfo  {journal} {Phys. Chem. Chem. Phys.}\ }\textbf {\bibinfo
  {volume} {4}},\ \bibinfo {pages} {4285--4291} (\bibinfo {year}
  {2002})}\BibitemShut {NoStop}%
\bibitem [{\citenamefont {Boys}(1960)}]{Boys1960}%
  \BibitemOpen
  \bibfield  {author} {\bibinfo {author} {\bibfnamefont {S.~F.}\ \bibnamefont
  {Boys}},\ }\bibfield  {title} {\enquote {\bibinfo {title} {{Construction of
  some molecular orbitals to be approximately invariant for changes from one
  molecule to another}},}\ }\href {https://doi.org/10.1103/RevModPhys.32.296}
  {\bibfield  {journal} {\bibinfo  {journal} {Rev. Mod. Phys.}\ }\textbf
  {\bibinfo {volume} {32}},\ \bibinfo {pages} {296--299} (\bibinfo {year}
  {1960})}\BibitemShut {NoStop}%
\bibitem [{\citenamefont {Werner}\ \emph {et~al.}(2018)\citenamefont {Werner},
  \citenamefont {Knowles}, \citenamefont {Knizia}, \citenamefont {Manby},
  \citenamefont {{Sch\"{u}tz}}, \citenamefont {Celani}, \citenamefont
  {Gy\"orffy}, \citenamefont {Kats}, \citenamefont {Korona}, \citenamefont
  {Lindh}, \citenamefont {Mitrushenkov}, \citenamefont {Rauhut}, \citenamefont
  {Shamasundar}, \citenamefont {Adler}, \citenamefont {Amos}, \citenamefont
  {Bennie}, \citenamefont {Bernhardsson}, \citenamefont {Berning},
  \citenamefont {Cooper}, \citenamefont {Deegan}, \citenamefont {Dobbyn},
  \citenamefont {Eckert}, \citenamefont {Goll}, \citenamefont {Hampel},
  \citenamefont {Hesselmann}, \citenamefont {Hetzer}, \citenamefont {Hrenar},
  \citenamefont {Jansen}, \citenamefont {K\"oppl}, \citenamefont {Lee},
  \citenamefont {Liu}, \citenamefont {Lloyd}, \citenamefont {Ma}, \citenamefont
  {Mata}, \citenamefont {May}, \citenamefont {McNicholas}, \citenamefont
  {Meyer}, \citenamefont {{Miller III}}, \citenamefont {Mura}, \citenamefont
  {Nicklass}, \citenamefont {O'Neill}, \citenamefont {Palmieri}, \citenamefont
  {Peng}, \citenamefont {Pfl\"uger}, \citenamefont {Pitzer}, \citenamefont
  {Reiher}, \citenamefont {Shiozaki}, \citenamefont {Stoll}, \citenamefont
  {Stone}, \citenamefont {Tarroni}, \citenamefont {Thorsteinsson},
  \citenamefont {Wang},\ and\ \citenamefont {Welborn}}]{MOLPRO}%
  \BibitemOpen
  \bibfield  {author} {\bibinfo {author} {\bibfnamefont {H.-J.}\ \bibnamefont
  {Werner}}, \bibinfo {author} {\bibfnamefont {P.~J.}\ \bibnamefont {Knowles}},
  \bibinfo {author} {\bibfnamefont {G.}~\bibnamefont {Knizia}}, \bibinfo
  {author} {\bibfnamefont {F.~R.}\ \bibnamefont {Manby}}, \bibinfo {author}
  {\bibfnamefont {M.}~\bibnamefont {{Sch\"{u}tz}}}, \bibinfo {author}
  {\bibfnamefont {P.}~\bibnamefont {Celani}}, \bibinfo {author} {\bibfnamefont
  {W.}~\bibnamefont {Gy\"orffy}}, \bibinfo {author} {\bibfnamefont
  {D.}~\bibnamefont {Kats}}, \bibinfo {author} {\bibfnamefont {T.}~\bibnamefont
  {Korona}}, \bibinfo {author} {\bibfnamefont {R.}~\bibnamefont {Lindh}},
  \bibinfo {author} {\bibfnamefont {A.}~\bibnamefont {Mitrushenkov}}, \bibinfo
  {author} {\bibfnamefont {G.}~\bibnamefont {Rauhut}}, \bibinfo {author}
  {\bibfnamefont {K.~R.}\ \bibnamefont {Shamasundar}}, \bibinfo {author}
  {\bibfnamefont {T.~B.}\ \bibnamefont {Adler}}, \bibinfo {author}
  {\bibfnamefont {R.~D.}\ \bibnamefont {Amos}}, \bibinfo {author}
  {\bibfnamefont {S.~J.}\ \bibnamefont {Bennie}}, \bibinfo {author}
  {\bibfnamefont {A.}~\bibnamefont {Bernhardsson}}, \bibinfo {author}
  {\bibfnamefont {A.}~\bibnamefont {Berning}}, \bibinfo {author} {\bibfnamefont
  {D.~L.}\ \bibnamefont {Cooper}}, \bibinfo {author} {\bibfnamefont {M.~J.~O.}\
  \bibnamefont {Deegan}}, \bibinfo {author} {\bibfnamefont {A.~J.}\
  \bibnamefont {Dobbyn}}, \bibinfo {author} {\bibfnamefont {F.}~\bibnamefont
  {Eckert}}, \bibinfo {author} {\bibfnamefont {E.}~\bibnamefont {Goll}},
  \bibinfo {author} {\bibfnamefont {C.}~\bibnamefont {Hampel}}, \bibinfo
  {author} {\bibfnamefont {A.}~\bibnamefont {Hesselmann}}, \bibinfo {author}
  {\bibfnamefont {G.}~\bibnamefont {Hetzer}}, \bibinfo {author} {\bibfnamefont
  {T.}~\bibnamefont {Hrenar}}, \bibinfo {author} {\bibfnamefont
  {G.}~\bibnamefont {Jansen}}, \bibinfo {author} {\bibfnamefont
  {C.}~\bibnamefont {K\"oppl}}, \bibinfo {author} {\bibfnamefont {S.~J.~R.}\
  \bibnamefont {Lee}}, \bibinfo {author} {\bibfnamefont {Y.}~\bibnamefont
  {Liu}}, \bibinfo {author} {\bibfnamefont {A.~W.}\ \bibnamefont {Lloyd}},
  \bibinfo {author} {\bibfnamefont {Q.}~\bibnamefont {Ma}}, \bibinfo {author}
  {\bibfnamefont {R.~A.}\ \bibnamefont {Mata}}, \bibinfo {author}
  {\bibfnamefont {A.~J.}\ \bibnamefont {May}}, \bibinfo {author} {\bibfnamefont
  {S.~J.}\ \bibnamefont {McNicholas}}, \bibinfo {author} {\bibfnamefont
  {W.}~\bibnamefont {Meyer}}, \bibinfo {author} {\bibfnamefont {T.~F.}\
  \bibnamefont {{Miller III}}}, \bibinfo {author} {\bibfnamefont {M.~E.}\
  \bibnamefont {Mura}}, \bibinfo {author} {\bibfnamefont {A.}~\bibnamefont
  {Nicklass}}, \bibinfo {author} {\bibfnamefont {D.~P.}\ \bibnamefont
  {O'Neill}}, \bibinfo {author} {\bibfnamefont {P.}~\bibnamefont {Palmieri}},
  \bibinfo {author} {\bibfnamefont {D.}~\bibnamefont {Peng}}, \bibinfo {author}
  {\bibfnamefont {K.}~\bibnamefont {Pfl\"uger}}, \bibinfo {author}
  {\bibfnamefont {R.}~\bibnamefont {Pitzer}}, \bibinfo {author} {\bibfnamefont
  {M.}~\bibnamefont {Reiher}}, \bibinfo {author} {\bibfnamefont
  {T.}~\bibnamefont {Shiozaki}}, \bibinfo {author} {\bibfnamefont
  {H.}~\bibnamefont {Stoll}}, \bibinfo {author} {\bibfnamefont {A.~J.}\
  \bibnamefont {Stone}}, \bibinfo {author} {\bibfnamefont {R.}~\bibnamefont
  {Tarroni}}, \bibinfo {author} {\bibfnamefont {T.}~\bibnamefont
  {Thorsteinsson}}, \bibinfo {author} {\bibfnamefont {M.}~\bibnamefont
  {Wang}},\ and\ \bibinfo {author} {\bibfnamefont {M.}~\bibnamefont
  {Welborn}},\ }\href@noop {} {\enquote {\bibinfo {title} {Molpro, version
  2018.3, a package of ab initio programs},}\ } (\bibinfo {year} {2018}),\
  \bibinfo {note} {see http://www.molpro.net}\BibitemShut {NoStop}%
\bibitem [{\citenamefont {Pulay}\ and\ \citenamefont
  {Saeb{\o}}(1986)}]{LMP2gradient}%
  \BibitemOpen
  \bibfield  {author} {\bibinfo {author} {\bibfnamefont {P.}~\bibnamefont
  {Pulay}}\ and\ \bibinfo {author} {\bibfnamefont {S.}~\bibnamefont
  {Saeb{\o}}},\ }\bibfield  {title} {\enquote {\bibinfo {title}
  {Orbital-invariant formulation and second-order gradient evaluation in
  m{\o}ller-plesset perturbation theory},}\ }\href@noop {} {\bibfield
  {journal} {\bibinfo  {journal} {Theor. Chim. Acta}\ }\textbf {\bibinfo
  {volume} {69}},\ \bibinfo {pages} {357--368} (\bibinfo {year}
  {1986})}\BibitemShut {NoStop}%
\bibitem [{\citenamefont {Pedregosa}\ \emph {et~al.}(2011)\citenamefont
  {Pedregosa}, \citenamefont {Varoquaux}, \citenamefont {Gramfort},
  \citenamefont {Michel}, \citenamefont {Thirion}, \citenamefont {Grisel},
  \citenamefont {Blondel}, \citenamefont {Prettenhofer}, \citenamefont {Weiss},
  \citenamefont {Dubourg}, \citenamefont {Vanderplas}, \citenamefont {Passos},
  \citenamefont {Cournapeau}, \citenamefont {Brucher}, \citenamefont {Perrot},\
  and\ \citenamefont {Duchesnay}}]{scikit-learn}%
  \BibitemOpen
  \bibfield  {author} {\bibinfo {author} {\bibfnamefont {F.}~\bibnamefont
  {Pedregosa}}, \bibinfo {author} {\bibfnamefont {G.}~\bibnamefont
  {Varoquaux}}, \bibinfo {author} {\bibfnamefont {A.}~\bibnamefont {Gramfort}},
  \bibinfo {author} {\bibfnamefont {V.}~\bibnamefont {Michel}}, \bibinfo
  {author} {\bibfnamefont {B.}~\bibnamefont {Thirion}}, \bibinfo {author}
  {\bibfnamefont {O.}~\bibnamefont {Grisel}}, \bibinfo {author} {\bibfnamefont
  {M.}~\bibnamefont {Blondel}}, \bibinfo {author} {\bibfnamefont
  {P.}~\bibnamefont {Prettenhofer}}, \bibinfo {author} {\bibfnamefont
  {R.}~\bibnamefont {Weiss}}, \bibinfo {author} {\bibfnamefont
  {V.}~\bibnamefont {Dubourg}}, \bibinfo {author} {\bibfnamefont
  {J.}~\bibnamefont {Vanderplas}}, \bibinfo {author} {\bibfnamefont
  {A.}~\bibnamefont {Passos}}, \bibinfo {author} {\bibfnamefont
  {D.}~\bibnamefont {Cournapeau}}, \bibinfo {author} {\bibfnamefont
  {M.}~\bibnamefont {Brucher}}, \bibinfo {author} {\bibfnamefont
  {M.}~\bibnamefont {Perrot}},\ and\ \bibinfo {author} {\bibfnamefont
  {E.}~\bibnamefont {Duchesnay}},\ }\bibfield  {title} {\enquote {\bibinfo
  {title} {{Scikit-learn: machine learning in python (v0.21.2)}},}\ }\href@noop
  {} {\bibfield  {journal} {\bibinfo  {journal} {J. Mach. Learn. Res.}\
  }\textbf {\bibinfo {volume} {12}},\ \bibinfo {pages} {2825} (\bibinfo {year}
  {2011})}\BibitemShut {NoStop}%
\bibitem [{\citenamefont {Okuta}\ \emph {et~al.}(2017)\citenamefont {Okuta},
  \citenamefont {Unno}, \citenamefont {Nishino}, \citenamefont {Hido},\ and\
  \citenamefont {Loomis}}]{cupy_learningsys2017}%
  \BibitemOpen
  \bibfield  {author} {\bibinfo {author} {\bibfnamefont {R.}~\bibnamefont
  {Okuta}}, \bibinfo {author} {\bibfnamefont {Y.}~\bibnamefont {Unno}},
  \bibinfo {author} {\bibfnamefont {D.}~\bibnamefont {Nishino}}, \bibinfo
  {author} {\bibfnamefont {S.}~\bibnamefont {Hido}},\ and\ \bibinfo {author}
  {\bibfnamefont {C.}~\bibnamefont {Loomis}},\ }\bibfield  {title} {\enquote
  {\bibinfo {title} {Cupy: A numpy-compatible library for nvidia gpu
  calculations},}\ }in\ \href
  {http://learningsys.org/nips17/assets/papers/paper_16.pdf} {\emph {\bibinfo
  {booktitle} {Proceedings of Workshop on Machine Learning Systems
  (LearningSys) in The Thirty-first Annual Conference on Neural Information
  Processing Systems (NIPS)}}}\ (\bibinfo {year} {2017})\BibitemShut {NoStop}%
\end{thebibliography}%


%
\end{document}